\documentclass[preprint,3p,12pt]{elsarticle}

\usepackage{mathrsfs}
\usepackage{amsmath}
\usepackage{stmaryrd}
\usepackage{bbding}
\usepackage{dcolumn}
\usepackage{graphicx}
\usepackage{amsfonts}
\usepackage{amssymb}
\usepackage{psfrag}
\usepackage{wrapfig}
\usepackage{makeidx}
\usepackage{bm}
\usepackage{epsf}
\usepackage{epsfig}
\usepackage{setspace}
\usepackage{epstopdf}
\usepackage{color}
\usepackage{comment}
\usepackage{algorithm}
\usepackage{algorithmic}
\usepackage{enumitem}
\usepackage{subcaption}
\newcommand \td {\mathrm{~d}}

\journal{}

\begin{document}

\begin{frontmatter}



\title{Unified Gas-Kinetic Scheme for Unsteady Multiscale Flows with Moving Boundaries}

\author[HKUST1]{Yue Zhang}
\ead{yzhangnl@connect.ust.hk}	
\author[xijiao]{Wenpei Long} 
\ead{wlongab@connect.ust.hk}
\author[HKUST1]{Junzhe Cao} 
\ead{jcaobb@connect.ust.hk}
\author[HKUST1,HKUST2,HKUST3]{Kun Xu\corref{cor1}}
\ead{makxu@ust.hk}

\address[HKUST1]{Department of Mathematics, Hong Kong University of Science and Technology, Clear Water Bay, Kowloon, Hong Kong}
\address[HKUST2]{Department of Mechanical and Aerospace Engineering, Hong Kong University of Science and Technology, Clear Water Bay, Kowloon, Hong Kong}
\address[HKUST3]{Shenzhen Research Institute, Hong Kong University of Science and Technology, Shenzhen, China}
\cortext[cor1]{Corresponding author}

\begin{abstract}
Simulating multiscale flows with moving boundaries, such as hypersonic multi-body separation and flows in micro-electro-mechanical systems (MEMS), requires robust numerical methods that couple mesh deformation with complex flow physics. This paper presents a hybrid overlapping moving-mesh technique developed within the unified gas-kinetic scheme (UGKS). To mitigate the Courant-Friedrichs-Lewy (CFL) constraint, we extend the implicit unsteady UGKS solver to support moving meshes, incorporating memory-efficient data handling and parallel computing optimizations to maximize computational efficiency. Validated against hypersonic multi-body separation and thermal rarefied MEMS flows, the proposed scheme accurately resolves complex, dynamic multiscale phenomena. The results confirm that this robust and efficient method provides a highly reliable tool for modeling dynamic flow interactions in complex geometric configurations.
\end{abstract}


\end{frontmatter}


\section{Introduction}

The development of near-space hypersonic vehicles faces a major bottleneck in multi-body separation, where transient and unsteady aerodynamic coupling severely impacts mission reliability. Similarly, the continued advancement of Micro-Electro-Mechanical Systems (MEMS) relies on the accurate characterization of moving components, such as actuators and sensors. A unifying challenge in both fields is that their operating environments involve complex multi-scale flow phenomena \cite{li2019gas,wang2022investigation}, encompassing both rarefied and continuum regimes. Therefore, it is essential to develop advanced numerical methods capable of simulating multi-scale flows with dynamic moving meshes.

Due to the significantly reduced gas density in near-space environments and the extremely small characteristic lengths in MEMS devices, non-equilibrium flow phenomena become highly pronounced. Accurately capturing these non-equilibrium physics requires additional degrees of freedom, which challenges the validity of the macroscopic Navier-Stokes equations and necessitates the use of gas kinetic theory. At the fundamental level, the Boltzmann equation governs these flows, resolving physics at the scales of the particle mean free path and mean collision time.

Traditionally, approaches to simulating non-equilibrium flows fall into two categories: stochastic and deterministic methods. Stochastic methods, most notably the direct simulation Monte Carlo (DSMC) method \cite{bird1998molecular}, employ representative particles to track the microscopic gas distribution function and simulate its evolution. By characterizing non-equilibrium physics via adaptive particles in a local velocity space, DSMC achieves high computational efficiency for high-speed rarefied flows. Furthermore, because particle free-streaming and collisions strictly adhere to conservation laws, the method is highly robust. However, its inherent stochastic nature introduces statistical noise. Consequently, for low-speed flow simulations, DSMC requires a prohibitively large number of particles or extensive statistical averaging to achieve an acceptable signal-to-noise ratio.

Conversely, deterministic approaches, such as the discrete velocity method (DVM) \cite{chu1965kinetic,yang1995rarefied}, solve the Boltzmann equation (or related kinetic models) based on a discrete velocity distribution function. Because the gas fluxes at cell interfaces are constructed within this deterministic framework, these methods yield accurate solutions entirely free of statistical noise. Nevertheless, discretizing the global velocity space incurs substantial memory consumption and computational costs, particularly for three-dimensional, high-speed flow simulations.

A common limitation of both traditional stochastic and deterministic methods is the use of operator splitting to separate free transport and collisions, which often introduces numerical dissipation proportional to the time step. Consequently, the spatial mesh size and time step are strictly constrained by the particle mean free path and mean collision time, respectively. To overcome these limitations, unified gas-kinetic methods have been developed in recent years. Prominent examples include the DVM-based unified gas-kinetic scheme (UGKS) \cite{xu2010unified,juan-chen_huang_unified_2012}, the discrete UGKS (DUGKS) \cite{guo2013dugks}, and the particle-based unified gas-kinetic wave-particle (UGKWP) method \cite{liu2020ugkwp,zhu2019ugkwp}. By utilizing the integral or finite-difference solution of the kinetic equation along characteristics, these schemes introduce the time step into the flux evaluation as a physical observation scale, effectively coupling particle collisions with free transport. As a result, the physical laws are directly modeled within the discretized space \cite{xu2015direct}, and the unified preserving (UP) properties of these schemes have been rigorously proven \cite{guo2023unified}. These methods function as highly accurate Boltzmann solvers in the rarefied regime and seamlessly recover the Navier-Stokes equations in the continuum regime, bypassing the stringent constraints imposed by microscopic scales. Building upon these advancements, this paper extends the unsteady UGKS solver to accommodate moving-mesh problems by integrating an overlapping-mesh technique.

Furthermore, to make the unified gas-kinetic scheme (UGKS) viable for large-scale engineering applications, this paper incorporates advanced acceleration and optimization techniques. Recently, substantial efforts have been dedicated to enhancing the computational performance and reducing the memory footprint of multi-scale algorithms like UGKS and DUGKS. A prominent memory-reduction strategy is the implementation of an unstructured discrete velocity space (DVS) \cite{titarev2017numerical,chen2019conserved}, which drastically reduces the number of discrete velocity points required for three-dimensional problems without sacrificing accuracy. Similarly, the adaptive UGKS (AUGKS) \cite{xiao2020velocity} utilizes an adaptive velocity space decomposition to efficiently capture local non-equilibrium phenomena. By employing a discrete distribution function in highly non-equilibrium regions and a continuous distribution function in near-continuum regimes, AUGKS significantly conserves memory. This adaptive framework has been further extended to three-dimensional flows involving rotational and vibrational non-equilibrium \cite{wei2024adaptive}, and analogous adaptive strategies have been successfully applied to DUGKS \cite{yang2023adaptive} and UGKWP \cite{wei2023adaptive,cao2025adaptive}.

In terms of temporal acceleration, the integration of implicit methods has proven highly effective. Implicit UGKS (IUGKS) formulations have been developed for both steady \cite{zhu2016implicit,xu2022ugks} and unsteady \cite{zhu2019implicit} flows. By coupling macroscopic predictions with microscopic implicit iterations, IUGKS achieves convergence rates one to three orders of magnitude faster than its explicit counterpart across all flow regimes. This implicit framework has also been expanded to simulate three-dimensional thermal non-equilibrium flows with internal degrees of freedom \cite{zhang2024conservative}. Additionally, the incorporation of multigrid methods has yielded an order-of-magnitude increase in computational speed \cite{zhu2017unified}. Recently, these advancements culminated in the development of the implicit adaptive UGKS (IAUGKS) \cite{long2024implicit}, which synthesizes both adaptive and implicit techniques. Another notable advancement is the general synthetic iterative scheme \cite{su2020can,su2020fast}, which couples macroscopic equations with the Boltzmann equation to dynamically adjust the velocity distribution function for rapid convergence. Building upon these methodological advancements, the present work extends the unsteady implicit UGKS \cite{zhu2019implicit} to accommodate moving-mesh problems. Furthermore, state-of-the-art memory-saving and parallelization-enhancement techniques \cite{zhang2025efficiency} are integrated into the implicit solver to maximize overall computational efficiency.

Parallel to algorithmic accelerations, the simulation of multi-scale flows involving moving meshes has garnered significant attention. Chen et al. \cite{chen2012unified} extended the UGKS to moving-mesh problems by employing a reference-frame method coupled with a velocity-space adaptation technique. Wang et al. \cite{wang2019arbitrary,wang2022investigation,wang2023arbitrary} developed an Arbitrary Lagrangian-Eulerian (ALE) based DUGKS to simulate transonic continuum and rarefied gas flows with moving boundaries. Similarly, Tao et al. \cite{tao2018combined,he2024thermal} combined the immersed boundary (IB) method with DUGKS to model particle-fluid interactions and thermal rarefied gas flows in MEMS devices. More recently, Zeng et al. \cite{zeng2025gsis} introduced a GSIS-ALE approach for moving-boundary problems in rarefied flows using an overlapping-mesh technique; by leveraging synthetic iterative schemes, this method significantly improves computational efficiency over explicit solvers and removes the Courant-Friedrichs-Lewy (CFL) timestep restriction. Building upon these foundations, the present work extends the unsteady implicit UGKS \cite{zhu2019implicit} to moving-mesh configurations via an overlapping-mesh technique, thereby achieving a robust, highly efficient solver free from CFL limitations.

The remainder of this paper is organized as follows. Section 2 introduces the implicit unified gas-kinetic scheme (UGKS) for unsteady flows. Section 3 details the extension of this implicit solver to moving-mesh problems utilizing an overlapping-mesh technique. Section 4 presents the numerical results to demonstrate the performance of the proposed method. Finally, Section 5 concludes the paper with a summary of the key findings.


\section{Implicit Unified Gas-Kinetic Scheme}
In this section, the implicit unified gas-kinetic scheme for unsteady flow \cite{zhu2019implicit} is introduced. In addition, mesh motion is accounted for in this scheme.
\subsection{Unified Gas-Kinetic Scheme}
In this work, the Shakhov model is used in the UGKS. With mesh motion, it can be written as

\begin{equation}
\label{BGK}
\frac{\partial f}{\partial t} + \mathbf{w}\cdot\nabla f =\frac{f^+-f}{\tau},
\end{equation}
where $f=f(\mathbf{x},t,\mathbf{v},\mathbf{\xi})$ is the distribution function for gas molecules at physical space location $\mathbf{x}$ with microscopic translation velocity $\mathbf{v}$ and internal velocity $\mathbf{\xi}$, $\mathbf{w}=\mathbf{v}-\mathbf{U}$ is the relative velocity between the gas molecules and the mesh velocity $\mathbf{U}$, $\tau$ is particle collision time, and $f^+$ is the modified equilibrium distribution function.
The modified equilibrium distribution function is given by

\begin{equation*}
f^+=g\left[ 1+(1-\text{Pr})\mathbf{c}\cdot\mathbf{q}\left(\frac{c^2}{\text{R}T}-5\right)/(5pRT)\right]=g+g^+,
\end{equation*}
where $g$ is the Maxwellian distribution, $\text{Pr}$ is the Prandtl number, $\mathbf{c}=\mathbf{v}-\mathbf{V}$ is the random velocity, $\mathbf{V}$ is the macroscopic velocity, $\mathbf{q}$ is the heat flux, $\text{R}$ is gas constant and $T$ is the temperature. The Maxwellian distribution is

\begin{equation*}
g=\rho {\left(\frac{\lambda}{\pi}\right)}^\frac{K+D}{2}e^{-\lambda ((\mathbf{v}-\mathbf{V})^2+\mathbf{\xi}^2)},
\end{equation*}
where $\rho$ is density, $\lambda=m/(2k_B T)$, $m$ is molecule mass, $k_B$ is Boltzmann constant, $D$ is the spatial dimension, $K$ is the number of internal degrees of freedom and $\mathbf{\xi}^2=\xi_1^2+\xi_2^2+\cdots+\xi_K^2$.

The collision terms satisfy the conservative constraint, or compatibility condition.

\begin{equation*}
\int (f^+-f)\Psi\td \mathbf{v}\td\mathbf{\xi}=0,
\end{equation*}
where $\Psi=(1,\mathbf{v},\frac{1}{2}(\mathbf{v}^2+\mathbf{\xi}^2))^T$ is the collision invariants and $\td\mathbf{\xi}=\td\xi_1\td\xi_2\cdots\td\xi_K$.
The macroscopic conservative variables and heat flux can be calculated via

\begin{equation*}
\begin{aligned}
\mathbf{W}&=\left(\begin{matrix}
\rho \\ \rho \mathbf{V} \\\rho E
\end{matrix}\right)=\int \Psi f\td\mathbf{v}\td\mathbf{\xi}, \
\mathbf{q}& = \frac{1}{2}\int(\mathbf{v}-\mathbf{V}) (|\mathbf{v}-\mathbf{V}|^2+\mathbf{\xi}^2)f\td\mathbf{v} \td \mathbf{\xi}.
\end{aligned}
\end{equation*}

To reduce memory usage, two reduced distribution functions $h$ and $b$ are introduced, defined by

\begin{equation*}
\begin{aligned}
h(\mathbf{x},t,\mathbf{v})&=\int f\td \mathbf{\xi}, \
b(\mathbf{x},t,\mathbf{v})&=\int \mathbf{\xi}^2f\td\mathbf{\xi},
\end{aligned}
\end{equation*}
By multiplying Eq.(\ref{BGK}) by $1$ and $\mathbf{\xi}^2$ and integrating over the inner degrees of freedom, the reduced BGK-Shakhov equation can be obtained as

\begin{equation}\label{BGKReduce}
\begin{aligned}
\frac{\partial h}{\partial t} + \mathbf{w}\cdot\nabla h =\frac{h^+-h}{\tau},\\
\frac{\partial b}{\partial t} + \mathbf{w}\cdot\nabla b =\frac{b^+-b}{\tau},
\end{aligned}
\end{equation}
where the reduced equilibrium distributions are

\begin{equation*}
  \begin{aligned}
 h^+&=H+H^+,\\
 b^+&=B+B^+.
  \end{aligned}
\end{equation*}
The corresponding reduced Maxwellian distribution $g$ becomes,

\begin{equation*}
  \begin{aligned}
H&=\int g\td\boldsymbol{\xi}=\rho\left(\frac{\lambda}{\pi}\right)^{D/2}e^{-\lambda(\boldsymbol{u}-\boldsymbol{U})^2},\\
B&=\int\boldsymbol{\xi}^2 g\td\boldsymbol{\xi}=\frac{3-D+K}{2\lambda}H,
  \end{aligned}
\end{equation*}
and the corresponding terms related to $g^+$ becomes

\begin{equation*}
  \begin{aligned}
 H^+&=\int g^+\td\boldsymbol{\xi} = \frac{4(1-Pr)\lambda^2}{5\rho}(\boldsymbol{u}-\boldsymbol{U})\cdot \boldsymbol{q}(2\lambda(\boldsymbol{u}-\boldsymbol{U})^2-2-D)H,\\
 B^+&=\int\boldsymbol{\xi}^2 g^+\td\boldsymbol{\xi} = \frac{4(1-Pr)\lambda^2}{5\rho}(\boldsymbol{u}-\boldsymbol{U})\cdot \boldsymbol{q}\{[ 2\lambda(\boldsymbol{u}-\boldsymbol{U})^2-D](3-D+K)-2K\}\frac{H}{2\lambda}.
  \end{aligned}
\end{equation*}
In the following introduction, $f$ is used instead of $h$ and $b$ for simplicity.

Similar with the whole physical domain $\Omega$ discretized into small cells $\Omega_i$, the whole velocity space is discretized into velocity space cells $\delta \mathbf{v}_k$. The governing equation of the distribution function at the discretized velocity points can be written as

\begin{equation}\label{BGKdis}
\frac{\partial f_{k}}{\partial t} + \mathbf{w}_k\cdot\nabla f_{k} =\frac{f^+_{k}-f_{k}}{\tau},
\end{equation}
where $f_{k}=f_{k}(\mathbf{x},t)=f(\mathbf{x},t,\mathbf{v}_k)$. By taking the integration of the above equation over the physical cell $\Omega_i$ and from time $t^n$ to $t^{n+1}$, the distribution function at the cell interface can be updated by

\begin{equation*}
f_{i,k}^{n+1}=f_{i,k}^{n} -\frac{1}{|\Omega_i|}\sum_{j\in N(i)} S_{ij}\mathcal{F}_{ij,k} + \int_{t^{n}}^{t^{n+1}} \frac{f^+_{i,k}-f_{i,k}}{\tau_i}\td t,
\end{equation*}
where $f_{i,k}^n$ and $f_{i,k}^{n+1}$ are the average of distribution function over cell $\Omega_i$ and $\delta \mathbf{v}_k$ at time $t^n$ and $t^{n+1}$, $|\Omega_i|$ denotes the volume of cell $i$, $N(i)$ is the set of all interface-adjcent neighboring cells of cell $i$ and $j$ is one of the neighboring cells of $i$.  The interface between cells $i$ and $j$ is labeled $ij$ and has area $S _ {ij}$. And the microscopic flux $\mathcal{F}_{ij,k}$ is

\begin{equation}\label{microscopicFlux}
\mathcal{F}_{ij,k} = w_{k,n}\int_{t^n}^{t^{n+1}} f_{ij,k}\td t,
\end{equation}
where $w_{k,n}=\mathbf{w}_k\cdot\mathbf{n}_{ij}$ is the normal velocity via surface $ij$ with normal direction $\mathbf{n}_{ij}$, and $f_{ij,k}$ is the average distribution fuction over cell $\delta \mathbf{v}_k$ at the center of cell interface $ij$.

By taking moments of the above equation and considering the compatibility condition, the governing equation of macroscopic conservative values can be obtained as

\begin{equation}\label{macroscopicUpdata}
\mathbf{W}_i^{n+1}=\mathbf{W}_i^n-\frac{1}{\Omega_i}\sum_{j\in N(i)} S_{ij} \mathbf{F}_{ij}  ,
\end{equation}
where $\mathbf{W}_i^n$ and $\mathbf{W}_{i}^{n+1}$ are cell-averaged conservative value at time $t^n$ and $t^{n+1}$, and the macroscopic flux is defined by

\begin{equation}\label{macroscopicFlux}
\mathbf{F}_{ij}=\sum_{k} \mathcal{V}_k\left(\begin{matrix}
\mathcal{H}_{ij,k} \\ \mathbf{v}_k \mathcal{H}_{ij,k} \\ \frac{1}{2}( \mathbf{v}_k^2\mathcal{H}_{ij,k}+\mathcal{B}_{ij,k})
\end{matrix}\right),
\end{equation}
where $ \mathcal{V}_k$ is the weight of velocity space cell $\delta \mathbf{v}_k$ (set as the volume of the cell for unstructured velocity space in this work), and $ \mathcal{H}_{ij,k}$ and $ \mathcal{B}_{ij,k}$ are defined by

\begin{equation*}
\begin{aligned}
\mathcal{H}_{ij,k} &= w_{k,n}\int_{t^n}^{t^{n+1}} h_{ij,k}\td t,\
\mathcal{B}_{ij,k} &= w_{k,n}\int_{t^n}^{t^{n+1}} b_{ij,k}\td t.
\end{aligned}
\end{equation*}

To construct the numerical fluxes at $\mathbf{x}_0=(0,0,0)^T$, the integral solution along the characteristic line $\mathbf{x}^\prime=\mathbf{x}_0-\mathbf{w}(t-t^\prime)$ of BGK equation (\ref{BGKdis}) gives

\begin{equation*}
f_{k}(\mathbf{x},t,\mathbf{\xi})=\frac{1}{\tau}\int_{t_0}^t f^+_{k}(\mathbf{x}^\prime,t^\prime)e^{-(t-t^\prime)/\tau}\td t^\prime + e^{-t/\tau}f_{0,k}(\mathbf{x}-\mathbf{w}_k t),
\end{equation*}
where $f_{0,k}(\mathbf{x})$ is the initial distribution function at the beginning of each step $t_n$, and $f^+(\mathbf{x},t)$ is the effective equilibrium state distributed in space and time around $\mathbf{x}$ and $t$. The integral solution provides a multiscale model of the evolution from an initial non-equilibrium distribution $f$ to an equilibrium distribution $f^+$ via collisions.

To achieve second-order accuracy, the initial distribution function $f_{0,k}(\mathbf{x})$ is approximated as

\begin{equation*}
f_{0,k}(\mathbf{x}) = \begin{cases}
f_{k}^l+\mathbf{x}\cdot \nabla f_{k}^l,&w_{k,n}>0,\\
f_{k}^r+\mathbf{x}\cdot \nabla  f_{k}^r,&w_{k,n}<0,\\
\end{cases}
\end{equation*}
where $f_{k}^l$ and $f_{k}^r$ are the reconstructed initial distribution functions at the left and right sides of the interface. The equilibrium state is approximated\cite{xu2021unified} as

\begin{equation*}
f^+_{k}(\mathbf{x},t) \approx g_{0,k}+ g^+_{0,k} + \mathbf{x}\cdot\nabla g_{0,k}+\frac{\partial g_{0,k}}{\partial t}t.
\end{equation*}
It is derived from the conservative variables contributed by all particles transported from both sides of the interface. Details of the distribution functions and derivatives are demonstrated in earlier work  \cite{xu2015direct,xu2010unified}.

In summary, the distribution function at the cell interface is

\begin{equation}\label{fullSolution}
f_{k}(\mathbf{0},t,\mathbf{\xi})=\begin{cases}
c_1f_{k}^l +c_2\mathbf{w}_k\cdot\nabla f_{k}^l + c_3(g_{0,k}+g^+_{0,k})+c_4\mathbf{w}_k\cdot\nabla g_{0,k}+c_5\partial_t  g_{0,k},& w_{k,n}>0, \\
c_1f_{k}^r +c_2\mathbf{w}_k\cdot\nabla f_{k}^r + c_3(g_{0,k}+g^+_{0,k})+c_4\mathbf{w}_k\cdot\nabla g_{0,k}+c_5\partial_t  g_{0,k},& w_{k,n}<0,
\end{cases}
\end{equation}
and

\begin{equation*}
\begin{aligned}
c_1 &= e^{-t/\tau}, \\
c_2 &=-t e^{-t/\tau}, \\
c_3 &=1- e^{-t/\tau}, \\
c_4 &= te^{-t/\tau}-\tau(1-e^{-t/\tau}), \\
c_5 &= t-\tau(1-e^{-t/\tau}). \\
\end{aligned}
\end{equation*}

In this paper, two types of boundary conditions are considered. The first type is artificially defined boundaries, such as inlets, outlets, and far fields. These boundaries are constructed using ghost cells. The conservation variables at the ghost cells use the same boundary conditions as those in the Euler equations. Here, the distribution function is set to a Maxwellian distribution corresponding to the conservative variables. When calculating the fluxes, the distribution functions at each cell interface are chosen based on the velocity direction at that interface.

The second type is real wall boundaries, which are treated with the Maxwellian isothermal boundary conditions. The microscopic distribution function at the wall face is defined as

\begin{equation*}
f_{w,k}=\begin{cases}
c_1f_{in,k}^l +c_2\mathbf{w}_k\cdot\nabla f_{in,k}^l + c_3(g_{in,k}+g^+_{in,k})+c_4\mathbf{w}_k\cdot\nabla g_{in,k}+c_5\partial_t  g_{in,k},& w_{k,n}>0 \\
g_w, & w_{k,n}<0
\end{cases},
\end{equation*}
where $w_{k,n}=\mathbf{w}_k\cdot \mathbf{n}_w$ is the projection of micro velocity on wall-normal direction $\mathbf{n}_w$, the $f_{in,k}$ and $\partial g_{in,k}$ are obtained by one-sided interpolation of microscopic distribution function and macroscopic values from the interior region, and $g_w$ is defined as

\begin{equation*}
g_w=\rho_w\left(\frac{m}{2\pi kT_w}\right)e^{-\frac{m((\mathbf{v}-\mathbf{v}_w)^2+\mathbf{\xi}^2)}{2kT_w}},
\end{equation*}
where $T_w$ and $\mathbf{v}_w$ is given wall temperature and velocity, and $\rho_w$ is calculated by

\begin{equation}
\label{maxWalldes}
\sum_{k}\mathcal{V}_k w_{k,n}\int_{t^n}^{t^{n+1}} h_{w,k}\td t=0,
\end{equation}
which means that no particles penetrate the wall.

Then, the flux of the microscopic distribution function can be obtained as

\begin{equation*}
\mathcal{F}_{w,k}=\begin{cases}
w_{k,n} ( c_1f_{in,k}^l +c_2\mathbf{w}_k\cdot\nabla f_{in,k}^l + c_3(g_{in,k}+g^+_{in,k})+c_4\mathbf{w}_k\cdot\nabla g_{in,k}+c_5\partial_t  g_{in,k}),& w_{k,n}>0 \\
w_{k,n}g_w, & w_{k,n}<0
\end{cases},
\end{equation*}
and the macroscopic flux $\mathbf{F}_w={(F_{w0},F_{w1},F_{w2},F_{w3},F_{w4})}^T$ can be obtained through Eq.(\ref{macroscopicFlux}). By denoting the surface flux of momentum equation as $\mathbf{F}_{mw}=(F_{w1},F_{w2},F_{w3})$, the surface pressure, shear stress, and heat flux can be obtained as

\begin{equation*}
p_w = \mathbf{F}_{mw}\cdot \mathbf{n}_w,
\sigma_w=|\mathbf{F}_{mw}-p_w\mathbf{n}_w|,
q_w=F_{w4}.
\end{equation*}

The source term can be discretized by the trapezoidal rule as

\begin{equation*}
\int_{t^{n}}^{t^{n+1}} \frac{f^+_{i,k}-f_{i,k}}{\tau_i}\td t = \frac{\Delta t}{2}\left(\frac{f^{+(n+1)}_{i,k}-f^{n+1}_{i,k}}{\tau^{n+1}_i} + \frac{f^{+(n)}_{i,k}-f^n_{i,k}}{\tau^n_i}\right),
\end{equation*}
where the $f^{+(n+1)}$ and $\tau^{n+1}$ can be obtained through the macroscopic conservative values $\mathbf{W}^{n+1}$.

\subsection{Implicit Unified Gas-Kinetic Scheme for Unsteady Flows}
In this section, the implicit unified gas-kinetic scheme with dual time-stepping is introduced. The macroscopic and microscopic governing equations can be writen as

\begin{equation*}
\begin{aligned}
\frac{\partial \mathbf{W}_i}{\partial t} &+ \frac{1}{|\Omega_i|}\sum_{j\in N(i)} S_{ij}\hat{\mathbf{F}}_{ij} = 0, \\
\frac{\partial f_{i,k}}{\partial t} &+ \frac{1}{|\Omega_i|}\sum_{j\in N(i)} S_{ij}\hat{\mathcal{F}}_{ij,k} = \frac{f^+_{i,k}-f_{i,k}}{\tau_i},
\end{aligned}
\end{equation*}
where $\hat{\mathbf{F}}_{ij}$ and $\hat{\mathcal{F}}_{ij,k}$ are the surface macroscopic flux and microscopic flux at the cell interface $ij$. To incorporate the multiscale properties of the UGKS into the above formulation, it is better to reconstruct the interface fluxes as time-averaged rather than instantaneous.  The time-averaged flux is defined as

\begin{equation*}
\hat{\mathcal{F}}_{ij,k}(t) =\frac{1}{\Delta t_s}\int{t}^{t+\Delta t_s} \mathcal{F}_{ij,k}(t)\td t,
\end{equation*}
where $\Delta t_s$ is the explicit time step with the CFL number 0.5, and the macroscopic flux $\hat{\mathbf{F}}{ij}$ is calculated by the microscopic flux $\hat{\mathcal{F}}_{ij,k}$ through Eq.(\ref{macroscopicFlux}).

The dual time-stepping is implemented by solving the following equation, with the physical time derivatives discretized using the Crank-Nicolson scheme. The full-discretization form of the macroscopic and microscopic governing equations is given as

\begin{align}
\label{dualTimeStepping_macro}
(\frac{1}{\Delta t^\prime}+\frac{1}{\Delta t})\Delta\mathbf{W}_i^{n+1,s} + \frac{1}{2|\Omega_i|}\sum_{j\in N(i)} S_{ij}\Delta\hat{\mathbf{F}}_{ij}^{n+1,s} &= \mathbf{R}_i^{n+1,s}, \\
\label{dualTimeStepping_micro}
\left(\frac{1}{\Delta t^\prime}+\frac{1}{\Delta t}+\frac{1}{2\tau^{n+1,s}_i}\right)\Delta f_{i,k}^{n+1,s} + \frac{1}{2|\Omega_i|}\sum_{j\in N(i)} S_{ij}\Delta\hat{\mathcal{F}}_{ij,k}^{n+1,s} &= r_{i,k}^{n+1,s},
\end{align}
with $\Delta\mathbf{W}_i^{n+1,s}=\mathbf{W}_i^{n+1,s+1}-\mathbf{W}_i^{n+1,s}$ and $\Delta f_{i,k}^{n+1,s}=f_{i,k}^{n+1,s+1}-f_{i,k}^{n+1,s}$. The $\Delta t^\prime$ is the virtual time step, which is calculated with CFL numeber 100 in this paper. The residual terms $\mathbf{R}_i^{n+1,s}$ and $r_{i,k}^{n+1,s}$ are defined as

\begin{equation*}
\begin{aligned}
\mathbf{R}_i^{n+1,s} &=- \frac{\mathbf{W}^{n+1,s}_i-\mathbf{W}^{n}_i}{\Delta t}-\frac{1}{2|\Omega_i|}\sum_{j\in N(i)} S_{ij}(\hat{\mathbf{F}}_{ij}^{n+1,s}+\hat{\mathbf{F}}_{ij}^{n}), \\
r_{i,k}^{n+1,s}&=- \frac{f^{n+1,s}_{i,k}-f^{n}_{i,k}}{\Delta t}-\frac{1}{2|\Omega_i|}\sum_{j\in N(i)} S_{ij}(\hat{\mathcal{F}}_{ij,k}^{n}+\hat{\mathcal{F}}_{ij,k}^{n+1,s})+\frac{f^{+,n+1,s}_{i,k}-f_{i,k}^{n+1,s}}{2\tau^{n+1,s}_i}+\frac{f^{+,n}_{i,k}-f_{i,k}^n}{2\tau^n_i}\\
&=-\frac{f^{n+1,s}_{i,k}-f^{n}_{i,k}}{\Delta t} + \frac{\hat{r}_{i,k}^n+\hat{r}_{i,k}^{n+1,s}}{2},
\end{aligned}
\end{equation*}
where the
\begin{equation}\label{residual_macro_n}
\hat{r}_{i,k}^n = -\frac{1}{|\Omega_i|}\sum_{j\in N(i)} S_{ij}\hat{\mathcal{F}}_{ij,k}^{n}+\frac{f^{+,n}_{i,k}-f_{i,k}^n}{\tau^n_i},
\end{equation}
and
\begin{equation}\label{residual_micro_n+1,s}
\hat{r}_{i,k}^{n+1,s} = -\frac{1}{|\Omega_i|}\sum_{j\in N(i)} S_{ij}\hat{\mathcal{F}}_{ij,k}^{n+1,s}+\frac{f^{+,n+1,s}_{i,k}-f_{i,k}^{n+1,s}}{\tau^{n+1,s}_i},
\end{equation}
are the residual terms at the $n$ and $n+1,s$ time steps, respectively.

For the terms on the left side of Eq.(\ref{dualTimeStepping_macro}), the Euler equation-based fluxes are adopted to simplify the variation of the macroscopic values

\begin{equation*}
\Delta\hat{\mathbf{F}}_{ij}^{n+1,s} \approx\frac{1}{2}\left[\Delta \mathbf{F}_{\text{Euler},i}^{n+1,s}+\Delta\mathbf{F}_{\text{Euler},j}^{n+1,s}+\Gamma_{ij}(\Delta\mathbf{W}_i^{n+1,s}-\Delta\mathbf{W}_j^{n+1,s})\right],
\end{equation*}
where the $\Delta$ variation of the Euler flux is defined as

\begin{equation*}
\Delta\mathbf{F}_{\text{Euler},i}^{n+1,s} = \mathbf{F}_{\text{Euler}}(\mathbf{W}^{n+1,s+1}_i)-\mathbf{F}_{\text{Euler}}(\mathbf{W}^{n+1,s}_i),
\end{equation*}
and the $\Gamma_{ij}=|(\mathbf{V}_{ij}-\mathbf{U}_{ij})\cdot\mathbf{n}_{ij}|+a$ is the spectral radius of the Euler flux Jacobian, where $a$ is the speed of sound. An upwind scheme is adopted to approximate the variation of the microscopic flux in Eq.(\ref{dualTimeStepping_micro}) as

\begin{equation*}
\Delta\hat{\mathcal{F}}_{ij,k}^{n+1,s} \approx\frac{1}{2}(1+\text{sign}(\hat{w}_{ij,k}))\hat{w}_{ij,k}\Delta f_{i,k}^{n+1,s}+\frac{1}{2}(1-\text{sign}(\hat{w}_{ij,k}))\hat{w}_{ij,k}\Delta f_{j,k}^{n+1,s},
\end{equation*}
where the $\hat{w}_{ij,k}=(\mathbf{v}_k-\mathbf{U}_{ij})\cdot\mathbf{n}_{ij}$ is the normal relative velocity of the cell interface $ij$.

For the current paper, the delta-form implicit scheme is solved by the lower-upper symmetric Gauss-Seidel (LU-SGS) scheme\cite{yoon1988lower}. The full algorithm for unsteady UGKS \cite{zhu2019implicit} are shown as follows:
\begin{enumerate}[label=\textbf{Step \arabic*} ]
\item Calculate the time interval $\Delta t_s$-averaged numerical flux $\hat{\mathbf{F}}_{ij}^{n}$ and $\hat{\mathcal{F}}_{ij,k}^{n}$ using the gas distribution function $f^{n}$ and macroscopic values $\mathbf{W}^{n}$.
\item Obtain the microscopic residual terms $\hat{r}_{i,k}^n$ using Eq.(\ref{residual_macro_n}) with $\mathbf{W}^{n}i$.
\item Compute the time interval $\Delta t_s$-averaged intermediate flux $\hat{\mathbf{F}}_{ij}^{n+1,s}$ and $\hat{\mathcal{F}}_{ij,k}^{n+1,s}$ using the gas distribution function $f^{n+1,s}$ and macroscopic values $\mathbf{W}^{n+1,s}$. For the first inner step, $f^{n+1,s}=f^{n}$ and $\mathbf{W}^{n+1,s}=\mathbf{W}^{n}$ are set.
\item Solve the delta-form implicit scheme of macroscopic values using Eq.(\ref{dualTimeStepping_macro}) by the LU-SGS scheme, and obtain the macroscopic values $\mathbf{W}^{n+1,s+1\prime}$.
\item Obtain the microscopic residual terms $\hat{r}_{i,k}^{n+1,s}$ using Eq.(\ref{residual_micro_n+1,s}) with $\mathbf{W}^{n+1,s+1\prime}_i$.
\item Solve the delta-form implicit scheme of microscopic values using Eq.(\ref{dualTimeStepping_micro}) by the LU-SGS scheme, and obtain the microscopic values $f^{n+1,s+1}$.
\item Fix the macroscopic values $\mathbf{W}^{n+1,s}$ using
\begin{equation*}
\mathbf{W}^{n+1,s+1}_i = \mathbf{W}^{n+1,s+1\prime}_i + \int(f^{n+1,s+1}_{i}-g^{n+1,s+1\prime}_{i})\Psi\td \mathbf{v}\td \mathbf{\xi},
\end{equation*}
where $f^{n+1,s+1}_{i}$ is the equilibrium state of the gas distribution function obtained by the conservative values $\mathbf{W}^{n+1,s+1}_i$.
\item Repeat the steps from 2 to 7 for the next inner step.
\end{enumerate}

\section{The algorithm of UGKS with overlapping mesh}

In the current paper, the UGKS with an overlapping mesh is implemented by the overset mesh assembly and interpolation. The overset mesh assembly is implemented using the grid zone division and a global-to-local parallel search strategy. Interpolation is implemented using the weighted average method. The whole algorithm is shown as follows:

\begin{enumerate}[label=\textbf{Step \arabic*} ]
  \item According to the Newton-Euler equation or given motion function, calculate the motion of the moving domain.
  \item Assemble the overset mesh.
  \item Evolve the solution according to the algorithm of unsteady UGKS introduced in Section 1.2.
  \item Go to step 1 for the next time step.
\end{enumerate}

\subsection{Overset mesh assembly}

To assemble the overset mesh, the grid zones are divided into subdomains, and each subdomain is assigned to a processor. The global-to-local parallel searching strategy introduced in \cite{chang2020parallel} is used to find donor cells in the nodes classified step and the donor cell finding step.

\begin{itemize}
  \item The preparatory work for each grid zone, such as building the alternating digital tree (ADT) data structure, calculating the minimum distance to the wall, and constructing some necessary geometric and topology information.
  \item For each node, find its donor cells in other grid zones. Classify nodes as active or inactive based on the minimum distance to the wall. Active nodes are those within the minimum distance to the wall, or the distance is less than the minimum distance to the wall in the moving domain.
  \item  Cells are categorized into three types: active, inactive, and boundary cells. A cell is active if all its nodes are active, inactive if all its nodes are inactive, and otherwise, it is a boundary cell. To achieve second-order accuracy, a buffer layer consisting of neighboring cells of boundary and inactive cells is added to the boundary cells.
  \item  Donor cells are found for each boundary cell and buffer layer cell in the overlap region.
\end{itemize}

\subsection{Interpolation}

To interpolate macroscopic variables and distribution functions, the interpolation for target cell $i$ is defined as follows:

\begin{equation*}
  \phi_i = \frac{\sum_{j\in N(i)} d_{ij}^{-2}\phi_j}{\sum_{j\in N(i)} d_{ij}^{-2}},
\end{equation*}

where $N(i)$ is the set of cells including the donor cell of the cell $i$ and its neighbor cells, $d_{ij}$ is the distance between two cells' centers, and $\phi$ is the macroscopic variable or distribution function. For the algorithm shown in Section 1.2, steps 1 and 3 are implemented in a memory-efficient manner \cite{zhang2025efficiency}. Cooperated with the overset interpolation, the whole algorithm is shown as Algorithm \ref{alg:fluxCalculation}, which shows that the interpolation is needed as the same as MPI communication.

\begin{algorithm}[!h]
 \caption{Parallel Algorithm for the UGKS flux calculation}
 \label{alg:fluxCalculation}
 \begin{algorithmic}[1]
  \STATE Calculate the time step $\Delta t_s$ based on the CFL condition and compute heat flux for each cell.
  \STATE Send macroscopic values, heat flux, and microscopic distribution function $f_1$.
  \STATE Receive macroscopic values, heat flux, and microscopic distribution function $f_1$.
  \STATE \textbf{Interpolate macroscopic variables, heat flux and distribution functions $f_1$ into the overlapping mesh.}
  \STATE Reconstruct macroscopic values.
  \STATE Reconstruct microscopic values for $f_1$ and store their gradients as $\nabla f_1$.
  \STATE Send gradients of macroscopic values, microscopic distribution function $f_2$, and gradients of microscopic values $\nabla f_{1}$.
  \FOR{$k=1$ to $k_{\text{max}}$}
    \STATE Let $\& \nabla f_{f} \gets (k \mod 2 == 1) \, ? \, \nabla f_{1} \, : \, \nabla f_{2}$ and $\&\nabla f_{r} \gets (k \mod 2 == 1) \, ? \, \nabla f_{2} \, : \, \nabla f_{1}$.
    \IF {$k==1$}
      \STATE Receive gradients of macroscopic values, distribution function $f_{2}$, and gradients of microscopic $\nabla f_{f}$.
      \STATE \textbf{Interpolate distribution functions $f_2$ into the overlapping mesh.}
      \STATE Calculate equilibrium state and time coefficients for cell interfaces.
    \ELSE
      \STATE Receive distribution function $f_{k+1}$ and gradients of microscopic $\nabla f_{f}$.
      \STATE \textbf{Interpolate distribution function $f_{k+1}$ into the overlapping mesh.}
    \ENDIF
    \STATE Reconstruct microscopic values for $f_{k+1}$ and store their gradients as $\nabla f_{r}$.
    \STATE Send microscopic distribution function $f_{k+2}$ and gradients of microscopic values $\nabla f_{r}$.
    \STATE Evolve flux $\mathcal{F}_{k}$ with $f_{k}$ and $\nabla f_{f}$ using Eq.(\ref{microscopicFlux}), and their contribution to macroscopic flux using Eq.(\ref{macroscopicFlux}).
    \STATE Interpolate $f_{k}$ and their gradients into the boundary and calculate the incident part of the flux.
  \ENDFOR
  \STATE Calculate the reflection part of the flux at the boundary.
 \end{algorithmic}
\end{algorithm}

Compared with the original algorithm in \cite{zhang2025efficiency}, several differences are introduced:

\begin{itemize}
  \item The residuals have to be stored due to the implicit scheme.
  \item About four times more memory than the original algorithm is needed to store the gas distribution function and its residuals because the dual time-stepping scheme is used.
  \item The wall boundary flux calculation was split into the incident part and the reflection part so that the gas distribution function and its gradients do not need to be stored in the boundary cells. Only two scalers, incident flux and reflection flux of density, are needed to be stored in the boundary cells.
\end{itemize}

\section{Numerical results}

\subsection{Forced motion of a micro beam in confined geometries}

\begin{figure}[!htbp]
  \centering
  \includegraphics[width=0.5\textwidth]{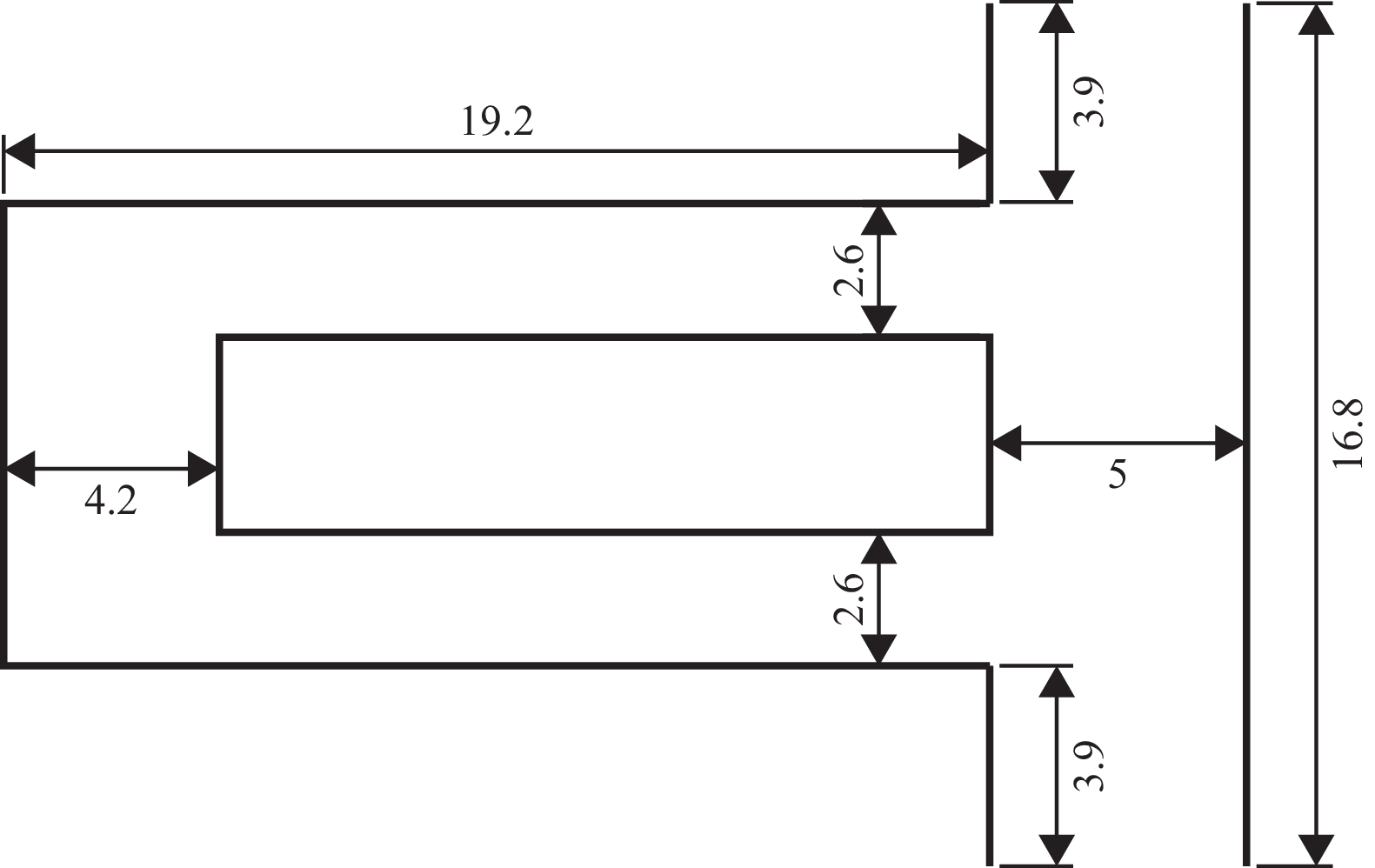}
  \caption{Schematic representation of the micro beam oscillation setup in confined space (unit: $\mathrm{\mu m}$)}\label{fig:microbeamGeom}
\end{figure}

In this numerical example, we investigate the rarefied gas flow within an open cavity containing a micro-beam structure. This configuration simulates the damping effects observed in MEMS comb drive mechanisms, particularly between the shuttle and stator components \cite{tsimpoukis2021linear}. The detailed configuration and dimensional specifications are presented in Fig. \ref{fig:microbeamGeom}.
The shuttle (micro beam) is positioned centrally in the left cavity at the initial state. The oscillatory motion of the shuttle, governed by $v_c=\cos(2\pi f_0t)$, induces fluid exchange through the upper and lower channels. Following the parameters established by Tiwari et al. \cite{tiwari2020interaction}, the oscillation operates at $f_0=176,000\text{Hz}$. The initial conditions specify a cavity pressure of $p_0=0.1\text{bar} =10,000\text{Pa}$, fluid temperature of $T_0=293\text{K}$, and density of $\rho_0=0.1641\text{kg/m}^3$. These parameters yield a Knudsen number of $\text{Kn}=0.2583$, referenced to the $4.2\mathrm{\mu m}$ gap width. The wall maintains a constant temperature of $T_w=293\text{K}$. The upper and lower channels implement free-flow conditions matching the initial state. The velocity space discretization spans $[-5\sqrt{RT_0},5\sqrt{RT_0}]^2$ with a uniform $49\times 49$ grid. The time step is set to $1\times 10^{-8}\text{s}$ and 10 inner steps are used for the dual time-stepping scheme.

\begin{figure}[!htbp]
  \centering
  \begin{subfigure}{0.4\textwidth}
    \includegraphics[width=0.9\textwidth]{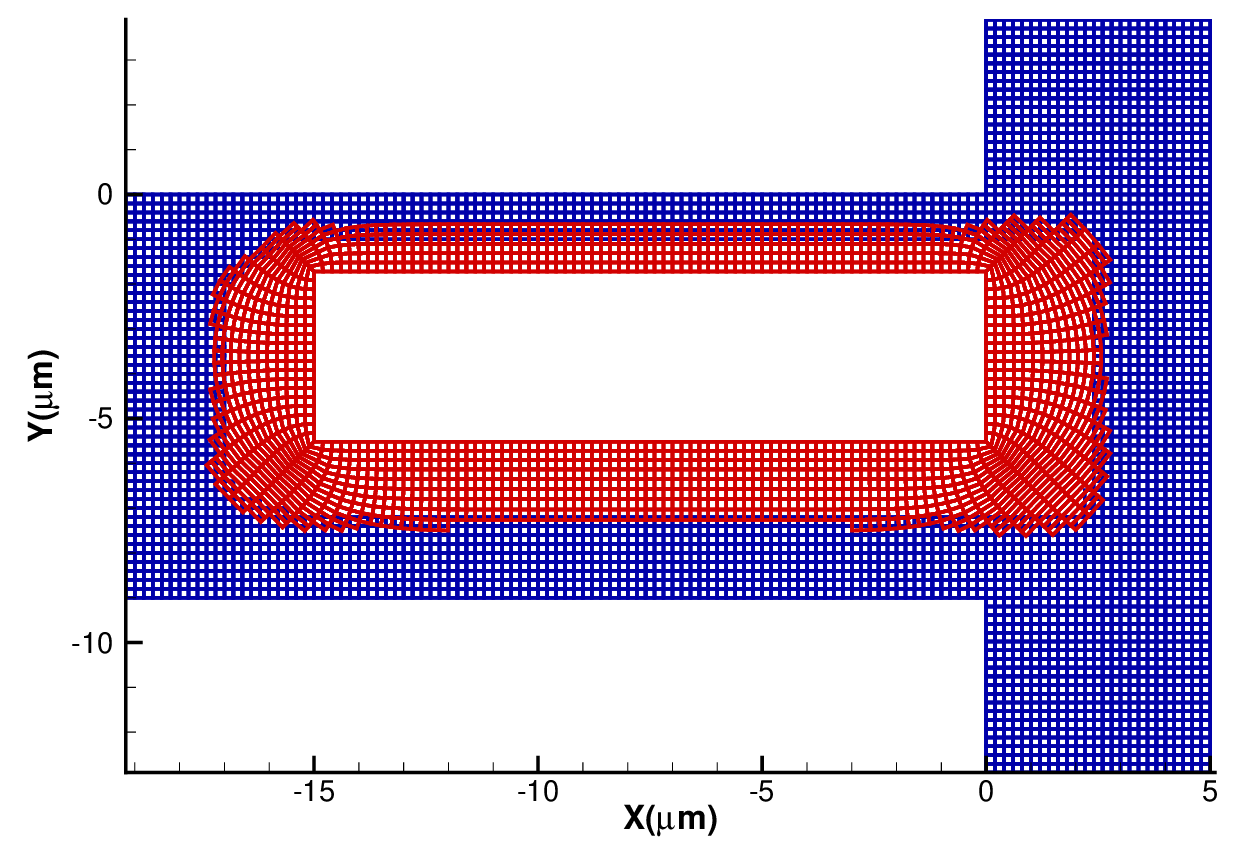}
    \caption{\label{meshMicroBeam}}
  \end{subfigure}
  \begin{subfigure}{0.4\textwidth}
    \includegraphics[width=0.9\textwidth]{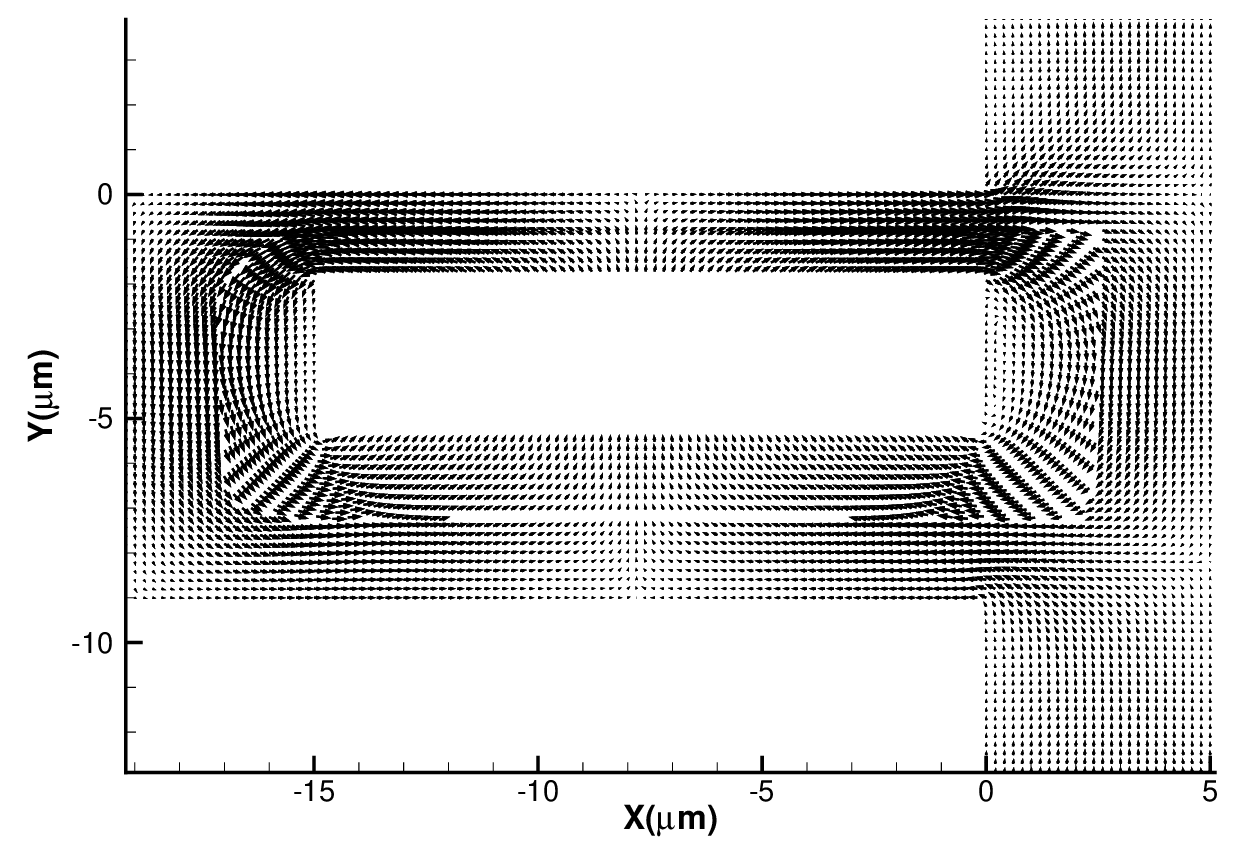}
    \caption{\label{microBeamvector12}}
  \end{subfigure}
  \begin{subfigure}{0.4\textwidth}
    \includegraphics[width=0.9\textwidth]{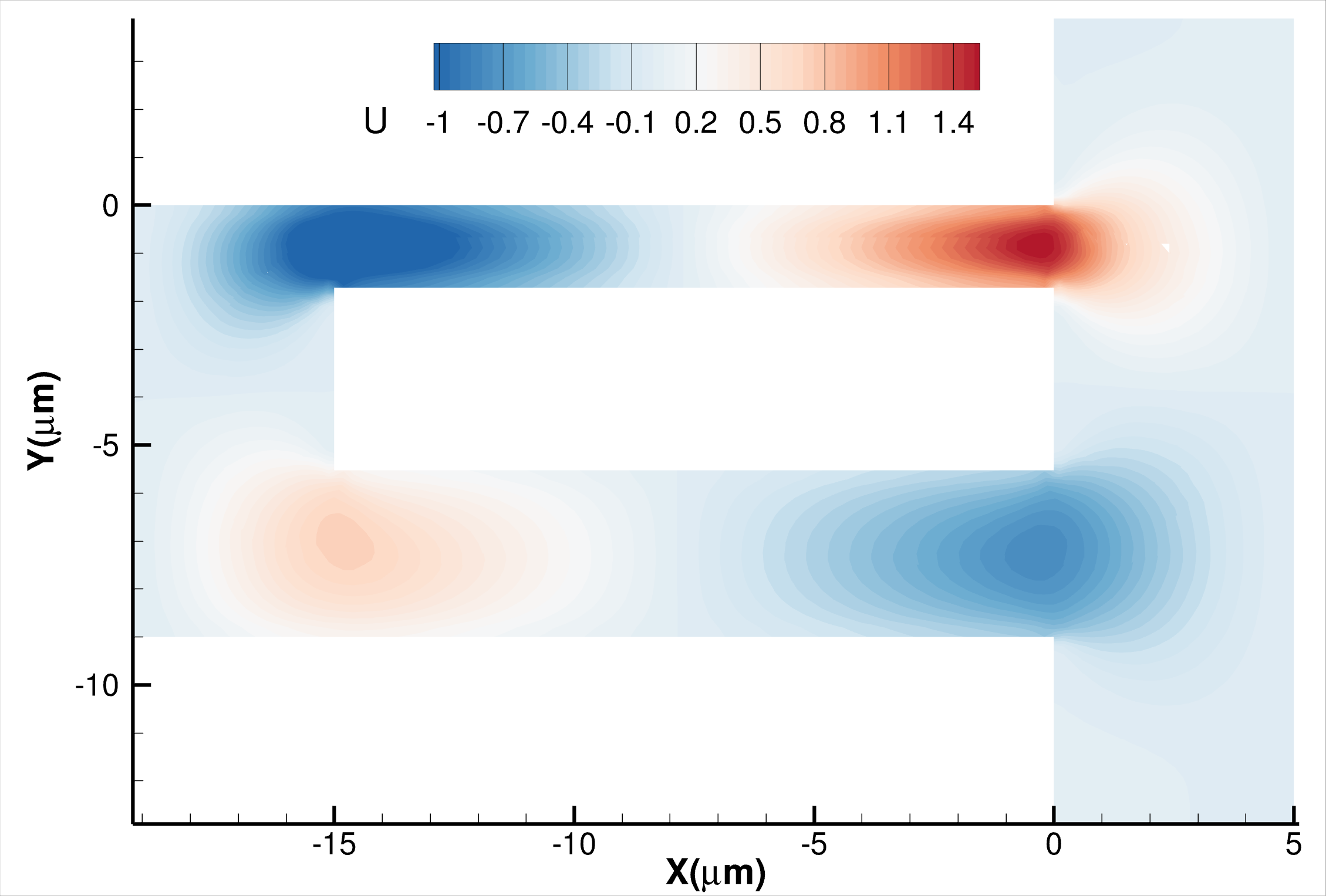}
    \caption{\label{microBeamU12}}
  \end{subfigure}
  \begin{subfigure}{0.4\textwidth}
    \includegraphics[width=0.9\textwidth]{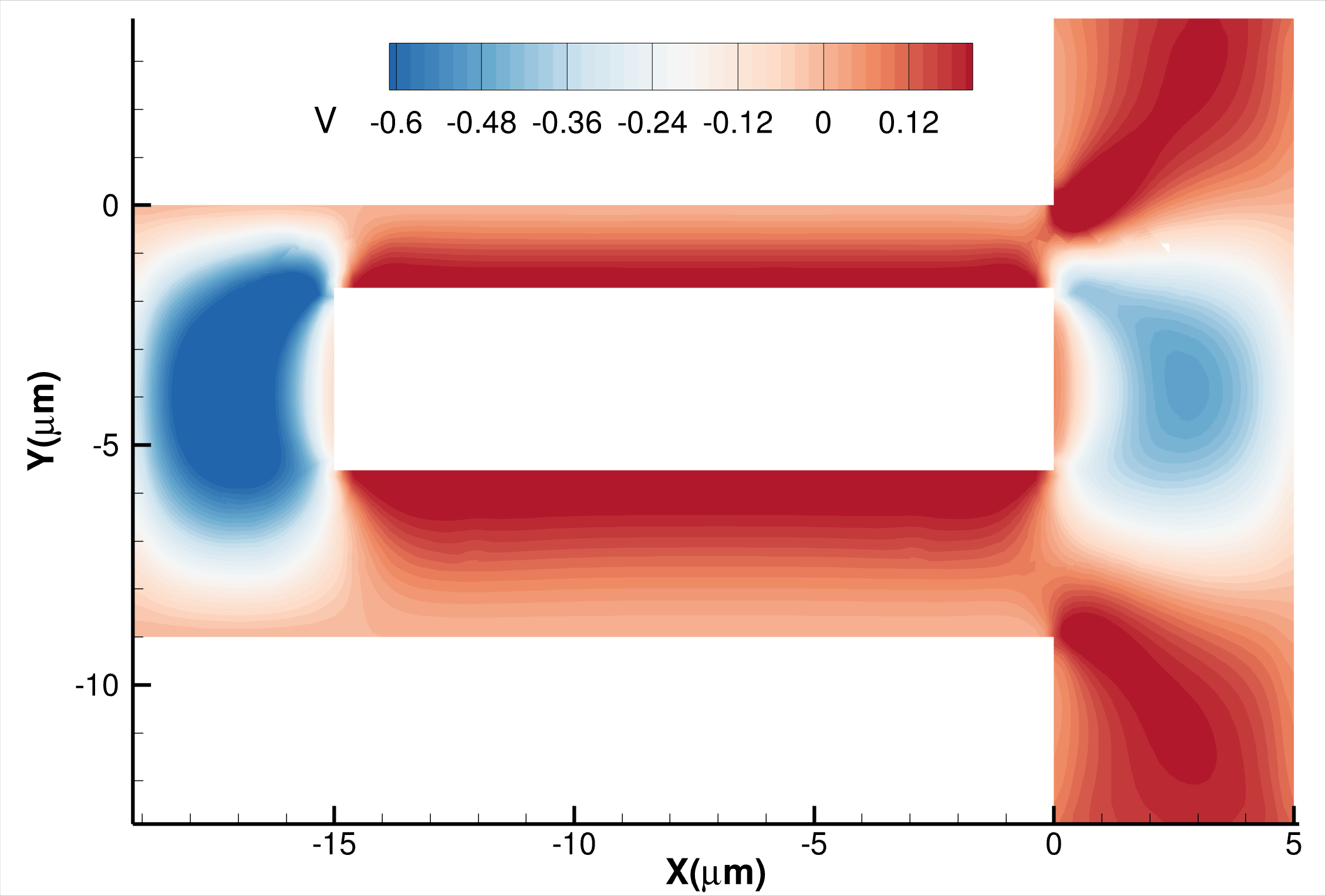}
    \caption{\label{microBeamV12}}
  \end{subfigure}
  \begin{subfigure}{0.4\textwidth}
    \includegraphics[width=0.9\textwidth]{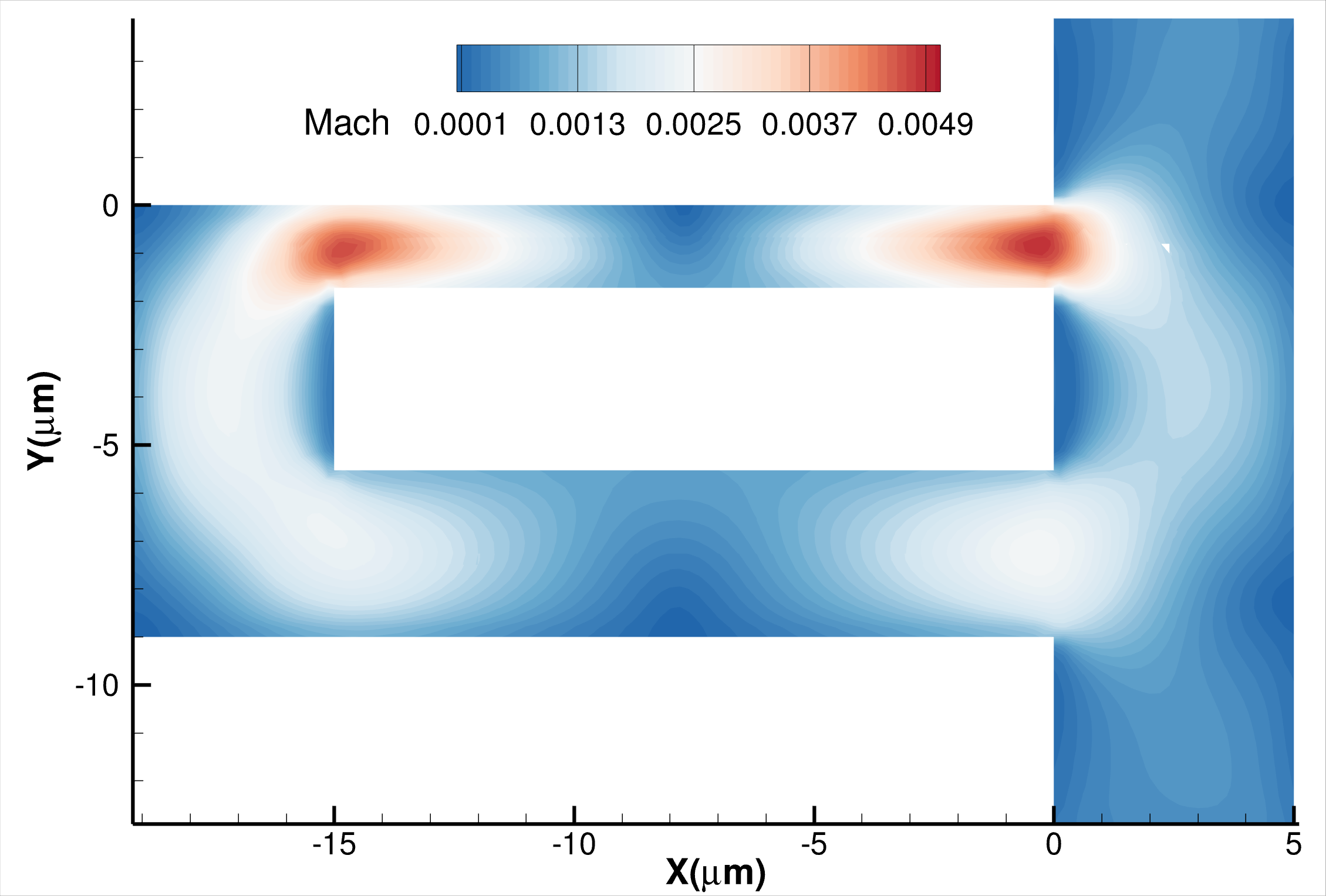}
    \caption{\label{microBeamMach12}}
  \end{subfigure}
  \begin{subfigure}{0.4\textwidth}
    \includegraphics[width=0.9\textwidth]{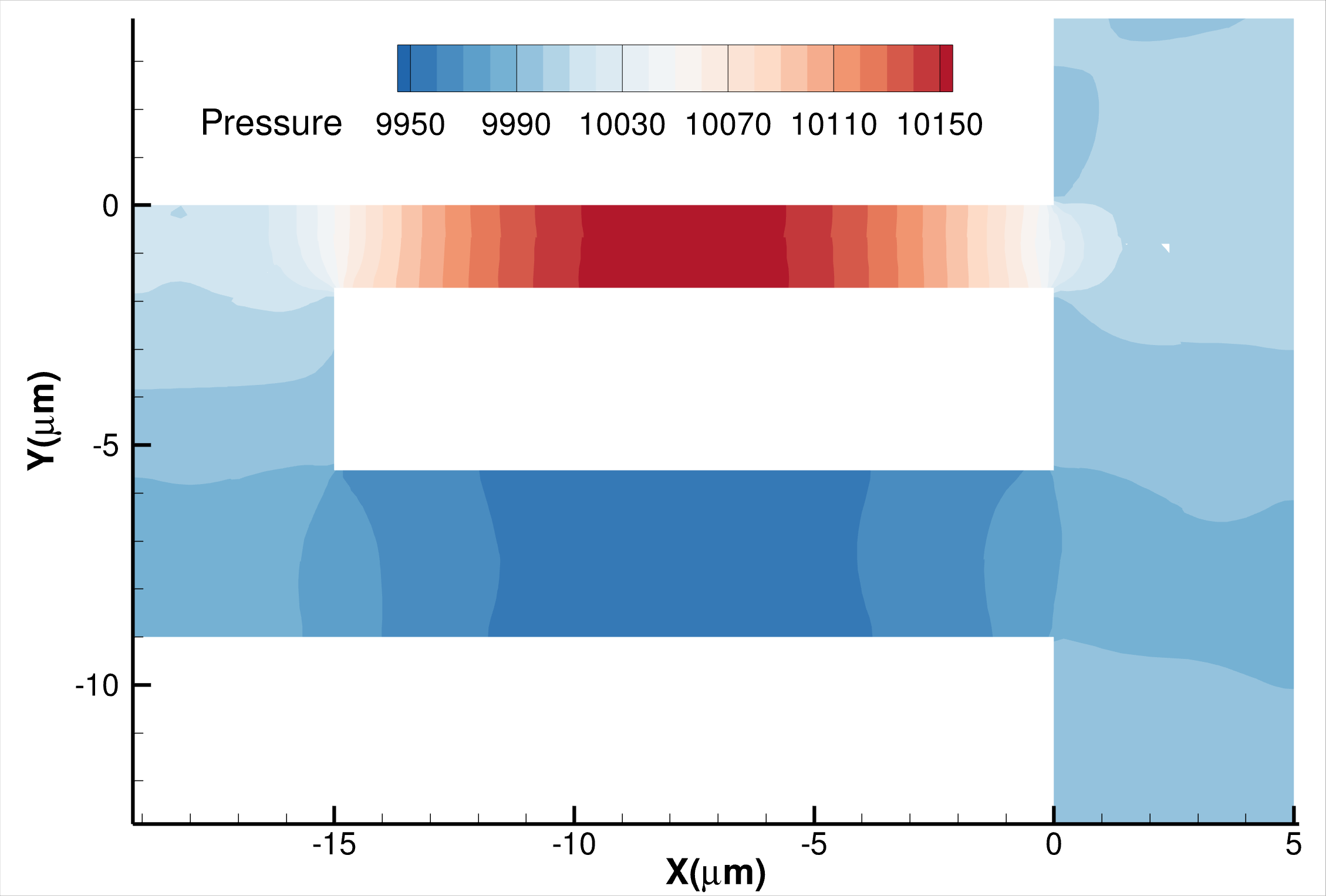}
    \caption{\label{microBeamPressure12}}
  \end{subfigure}
  \caption{Mesh distribution, velocity vector and contours at $t=1.2\times 10^{-6}\text{s}$ in the forced motion of a micro beam in confined geometries (a) Mesh assembly result, (b) The velocity vector, (c) The $x$-direction velocity contour, (d) The $y$-direction velocity contour, (e) The Mach number contour, (f) The Pressure contour}\label{fig:microbeam12}
\end{figure}

The computational results at $t=1.2\times 10^{-6}\text{s}$ are displayed in Fig.\ref{fig:microbeam12}. Fig.\ref{meshMicroBeam} shows the computational mesh structure, where red elements correspond to the micro beam region and black elements represent the background fluid domain. The interface between these regions exhibits smooth transitions. Fig.\ref{microBeamU12} reveals the $x$-velocity distribution, showing opposing flow patterns at the corners: positive velocities at top-right and lower-left, contrasting with negative velocities at top-left and lower-right. The upward shuttle motion compresses the upper cavity region, resulting in higher velocity magnitudes. This compression effect is further evidenced by the increased pressure and Mach number distributions shown in Fig.\ref{microBeamPressure12} and Fig.\ref{microBeamMach12}. The velocity vector field in Fig.\ref{microBeamvector12} highlights enhanced flow speeds near the top corners. The $y$-velocity contours demonstrate upward fluid motion along the shuttle's vertical walls, while downward flow occurs along the lateral sides. These flow characteristics align with the findings reported by Tiwari et al. \cite{tiwari2020interaction}.

\subsection{Free motion of a micro particle in a lid-driven cavity}

\begin{figure}[h]
    \centering
    \begin{subfigure}{0.4\textwidth}
        \includegraphics[width=0.9\textwidth]{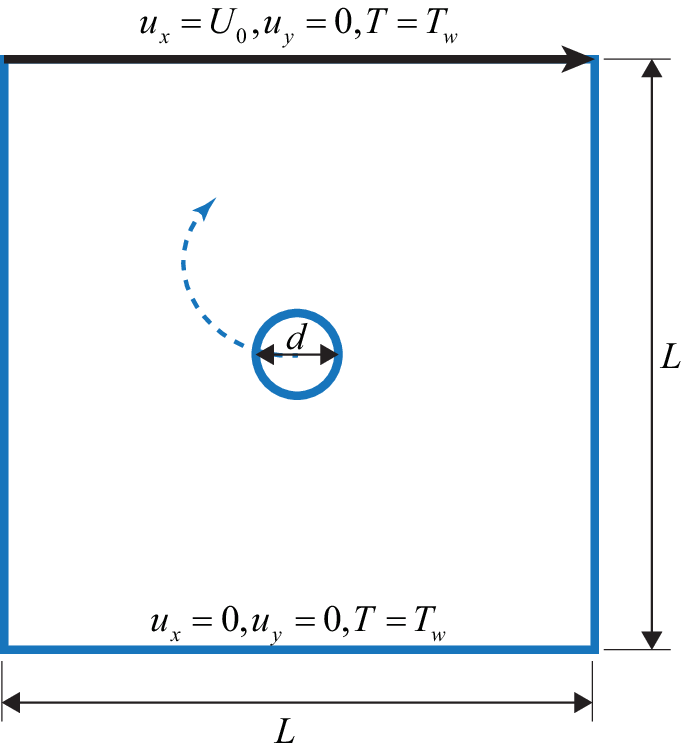}
        \caption{\label{fig:cavityGeom}}
    \end{subfigure}
    \begin{subfigure}{0.4\textwidth}
        \includegraphics[width=0.9\textwidth]{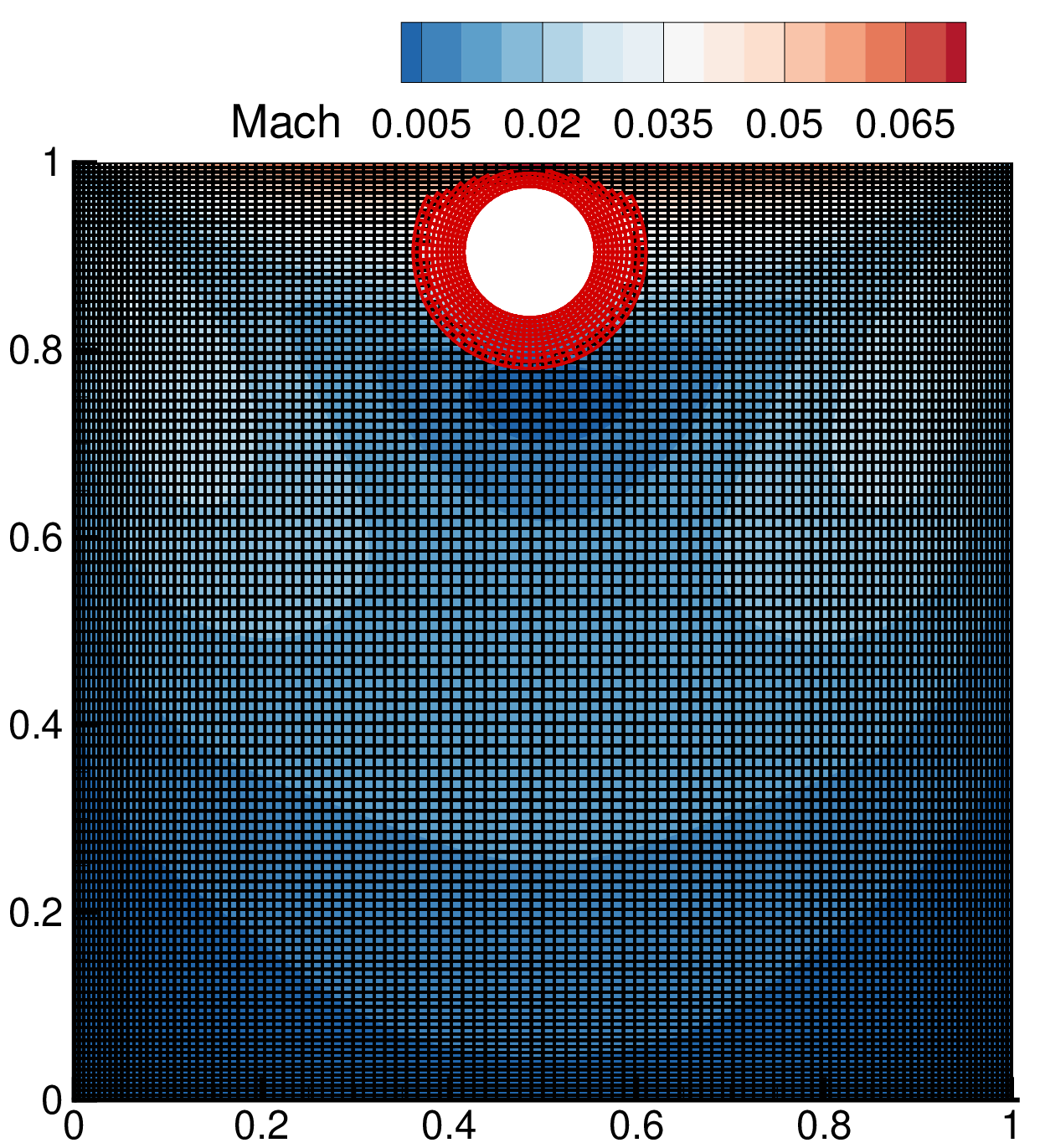}
        \caption{\label{fig:cavityMesh50}}
    \end{subfigure}
    \caption{The geometry, boundary conditions and mesh of the free motion of a micro particle in a lid-driven cavity (a) The geometry and boundary conditions (b) The mesh and Mach number contours at $t^*=50$}
    \label{fig:2}
\end{figure}

The investigation of particle-laden rarefied gas flows represents a significant research area, with the particle motion in a lid-driven cavity serving as a fundamental benchmark problem \cite{tiwari2020interaction,tao2018combined,zeng2025gsis}. The computational setup, illustrated in Fig.\ref{fig:cavityGeom}, consists of a square cavity with side length $L$ containing a spherical particle of diameter $d=0.14L$. The particle, having a density ratio of $\rho_c/\rho_f=10$ relative to the surrounding gas, is initially positioned at the cavity's center. The flow is driven by the top wall moving horizontally at a constant speed of $U_0=30\text{m/s}$, while the particle is allowed to move freely under the influence of fluid forces. The simulation parameters are configured with a fluid density of $\rho_f=1.103947\times 10^{-6}\text{kg/m}^3$, uniform temperature $T_w=T_f=273\text{K}$, and cavity length $L=1\text{m}$, yielding a Knudsen number of $0.1$. The computational domain employs a velocity-space discretization with $41\times 41$ points spanning $[-5\sqrt{RT_f},5\sqrt{RT_f}]^2$. The spatial domain utilizes $108\times 108$ elements for the background mesh, with $88$ uniform elements along the particle circumference. A refined mesh spacing of $0.005$ is implemented at the cavity boundaries and particle surface. The time step is set to $2\times 10^{-4}\text{s}$ and 10 inner steps are used for the dual time-stepping scheme.  The testing platform consists of an AMD Ryzen Threadripper 2990WX CPU operating at 4.2 GHz, featuring 32 cores and 128 GB of memory. The code is compiled with GCC 11.4.0 using the \texttt{-O2} optimization flag and linked with Open MPI 4.1.5. About 700 minutes are required for the entire simulation, with 2,300 physical time steps, on 32 cores.

\begin{figure}[h]
    \centering
    \begin{subfigure}{0.4\textwidth}
        \includegraphics[width=0.9\textwidth]{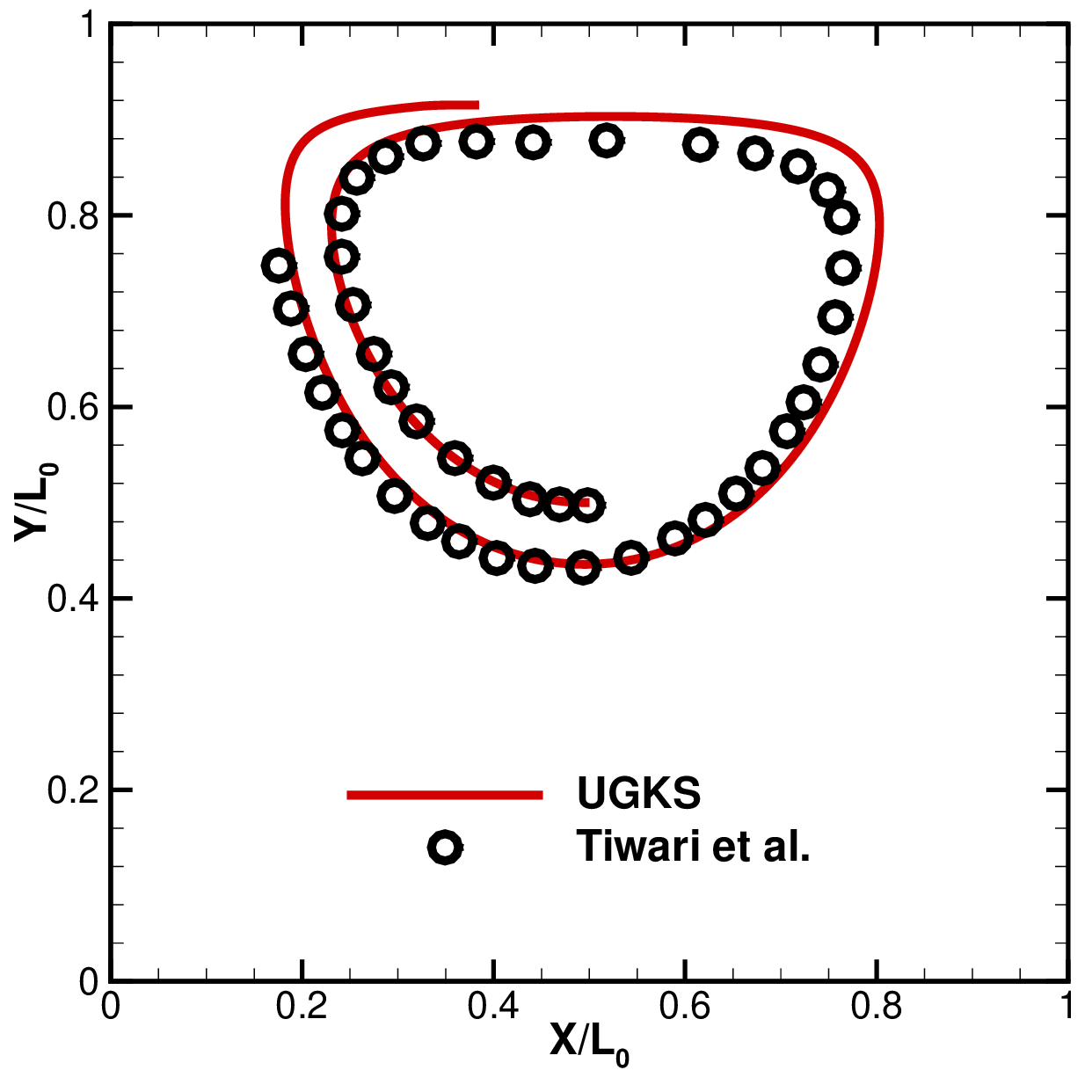}
        \caption{\label{fig:cavityTrace}}
    \end{subfigure}
    \begin{subfigure}{0.4\textwidth}
        \includegraphics[width=0.9\textwidth]{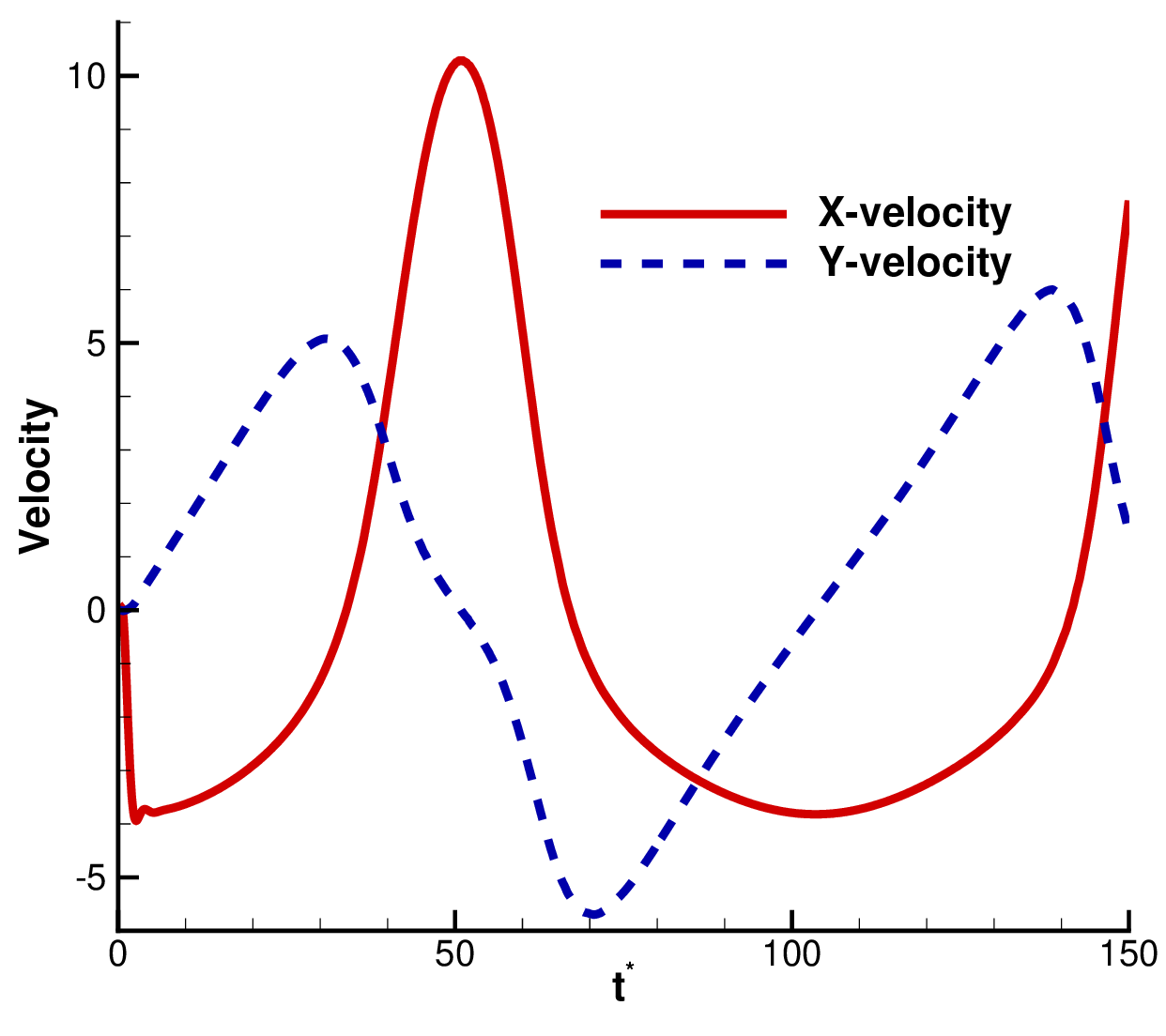}
        \caption{\label{fig:cavityVelocity}}
    \end{subfigure}
  \caption{The trace and the velocity history of the micro particle center (a) The trace of the micro particle center. (b) The velocity history of the micro particle center with the time normalized by $L/\sqrt{2RT_f}T$}
    \label{}
\end{figure}

The numerical results at dimensionless time $t^*=t\sqrt{2RT_f}/L=50$ are presented in Fig.\ref{fig:cavityMesh50}, depicting the mesh configuration and Mach number distribution as the particle approaches the cavity's upper region. The particle's trajectory, plotted in Fig.\ref{fig:cavityTrace}, demonstrates good agreement with previous numerical studies \cite{tiwari2020interaction}. The flow physics reveals an interesting pattern: the shearing action of the moving lid, combined with the confinement effect of the stationary walls, generates a circulating flow pattern that entrains the particle. Given its initial central position, the particle follows a clockwise path but exhibits a gradual outward drift rather than returning to its starting location. The particle's velocity components, shown in Fig.\ref{fig:cavityVelocity}, display distinct characteristics. Initially, there is a sharp acceleration in the $x$-direction as the particle is propelled leftward by the fluid motion. This motion is subsequently dampened by the left wall, leading to a reversal in direction. The trajectory analysis indicates that the particle migrates toward the upper cavity region while maintaining prolonged rightward motion. The presence of the right wall eventually redirects the particle toward the lower-left region, resulting in another reversal in the $x$-velocity. The vertical velocity component exhibits a characteristic rise-and-fall pattern throughout this motion.

\subsection{Two-stage-to-orbit (TSTO) hypersonic vehicles}

\begin{figure}[!htbp]
  \centering
  \begin{subfigure}{0.3\textwidth}
    \includegraphics[width=0.9\textwidth]{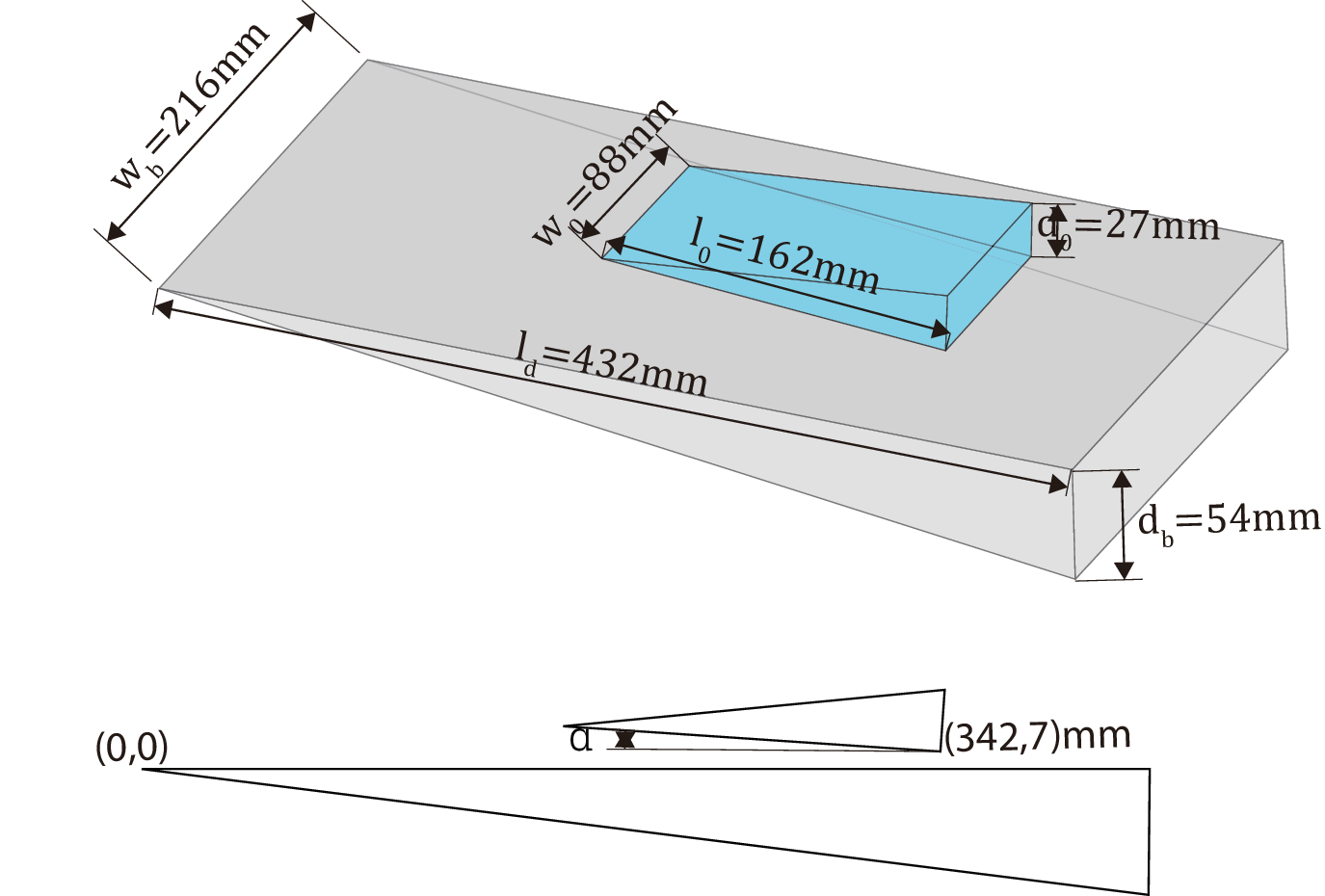}
    \caption{\label{fig:tstomodel}}
  \end{subfigure}
  \begin{subfigure}{0.3\textwidth}
    \includegraphics[width=0.9\textwidth]{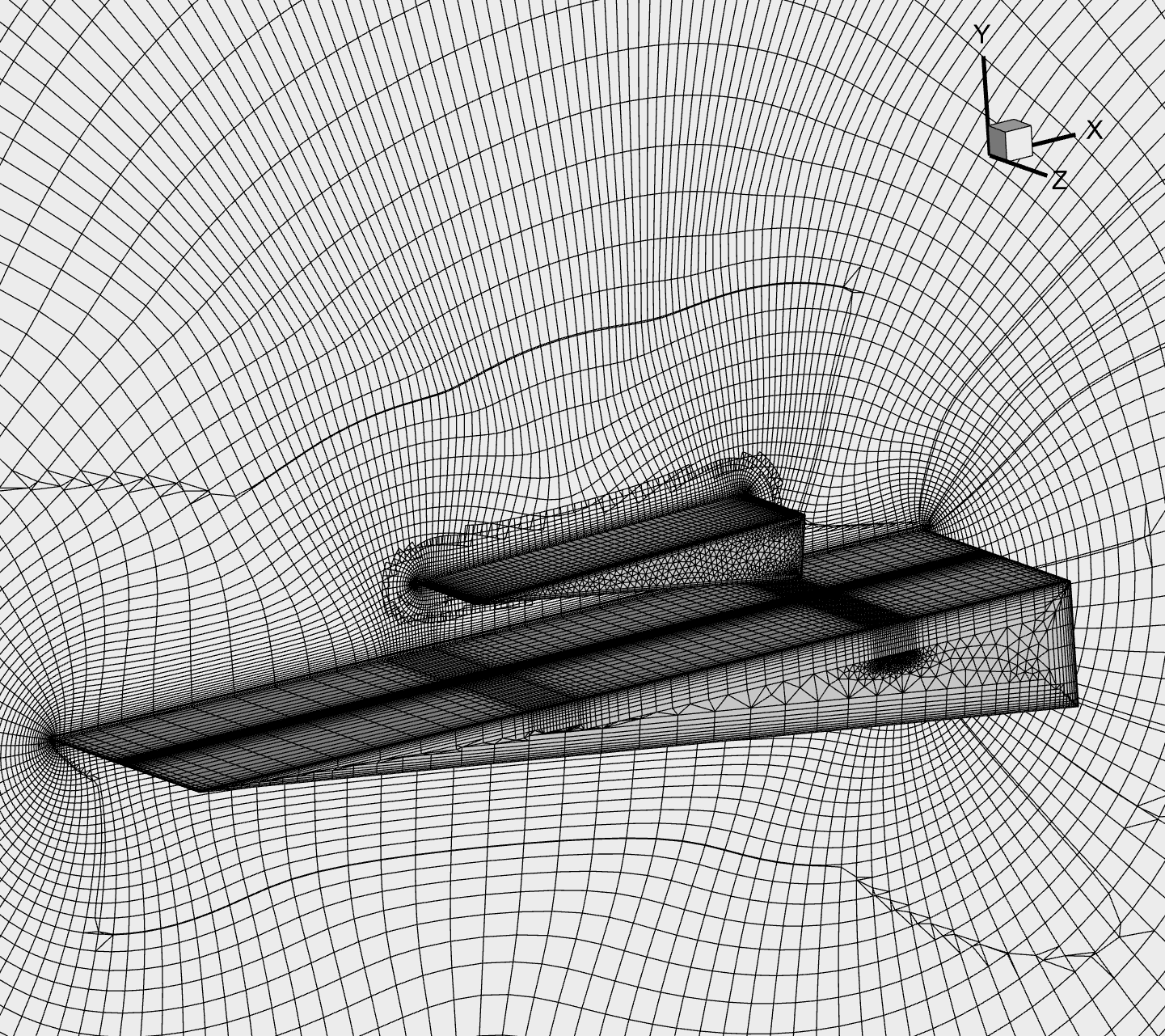}
    \caption{\label{fig:tstoMesh}}
  \end{subfigure}
  \begin{subfigure}{0.3\textwidth}
    \includegraphics[width=0.9\textwidth]{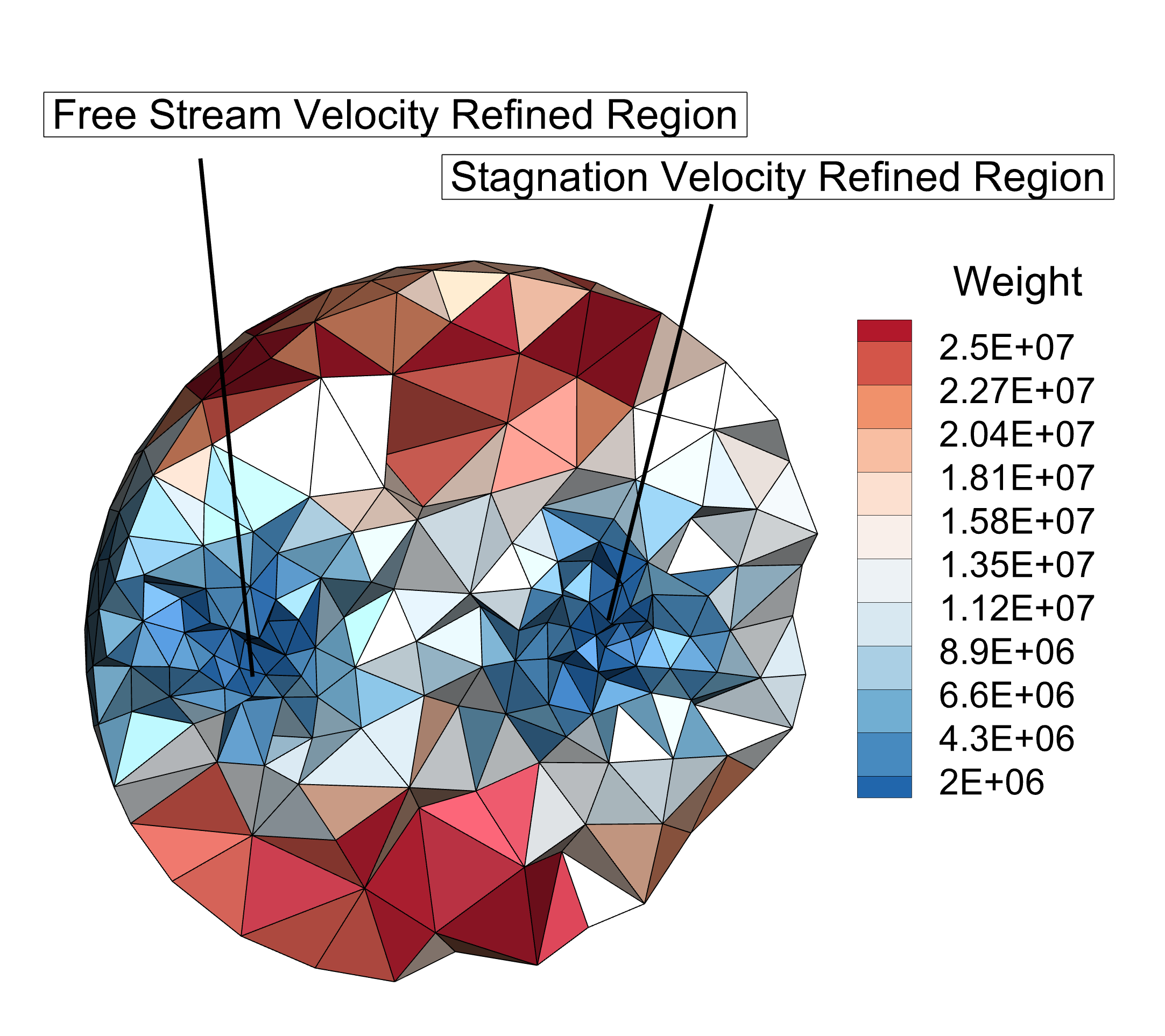}
    \caption{\label{fig:tstoDVS}}
  \end{subfigure}
  \caption{The discretization of physical space and velocity space. The geometry(a)  and initial mesh assembly (b) of the two-stage-to-orbit (TSTO) hypersonic vehicles. (c) The DVS mesh consisting of 4,748 cells}
\end{figure}

This model is used to test the three-dimensional mesh motion and the overset mesh assembly. The two-stage-to-orbit (TSTO) hypersonic vehicles are described in \cite{wang2022numerical}. The geometry is shown in Fig.\ref{fig:tstomodel}. The transverse stage separation (TSS) scheme, in which the orbiter moves along the direction normal to the booster's upper surface, is studied in this paper. The initial angle of incidence is set as $\alpha=8^\circ$. The incoming Mach number is $8$, and the Knudsen number is $0.1$ according to the reference length $l_o=0.162 \text{m}$. Argon is used as the working gas. The incoming flow is uniform with temperature $T_\infty=273 \text{K}$, density $\rho_\infty=5.3002\times10^{-6} \text{kg/m}^3$ and velocity $U_\infty=2461.917 \text{m/s}$. The mass of the orbiter is $m_o/\rho_\infty \cdot l_o^3=10$, and the moment of inertia are $I_{xx}/\rho_\infty \cdot l_o^5=6$, $I_{yy}/\rho_\infty \cdot l_o^5=12.5$, and $I_{zz}/\rho_\infty \cdot l_o^5=10$, respectively. The Newton-Euler equation is used to solve the motion of the orbiter. The time step is set to $5\times 10^{-6} \text{s}$ and 15 inner steps are used for the dual time-stepping scheme. At the initial state, the mesh is shown in Fig.\ref{fig:tstoMesh}. Totally, $1,258,788$ elements are used for the background mesh, and $264,570$ for the orbit mesh. The height of the first layer of the mesh is $5\times 10^{-4} \text{m}$. The DVS is discretized into $4,748$ cells in a sphere mesh with center at $(0.4U_\infty,0,0)$ a radius of $8\sqrt{\gamma RT_\infty}$. The velocity space near the zero and free stream velocity points is refined within a radius of $\sqrt{RT_\infty}$. The simulation was conducted on the SUGON computation platform using dual Hygon 7285H processors, each equipped with 32 cores, for a total of 64 cores at a base frequency of 2.5 GHz. The system architecture leverages the C86-2G microarchitecture, and the computation network is supported by a 100Gb InfiniBand (IB) connection. 256 cores are used for this computation, and about 4,615 seconds are required for one physical time step.

\begin{figure}[!htbp]
  \centering
  \begin{subfigure}{0.3\textwidth}
    \includegraphics[width=0.9\textwidth]{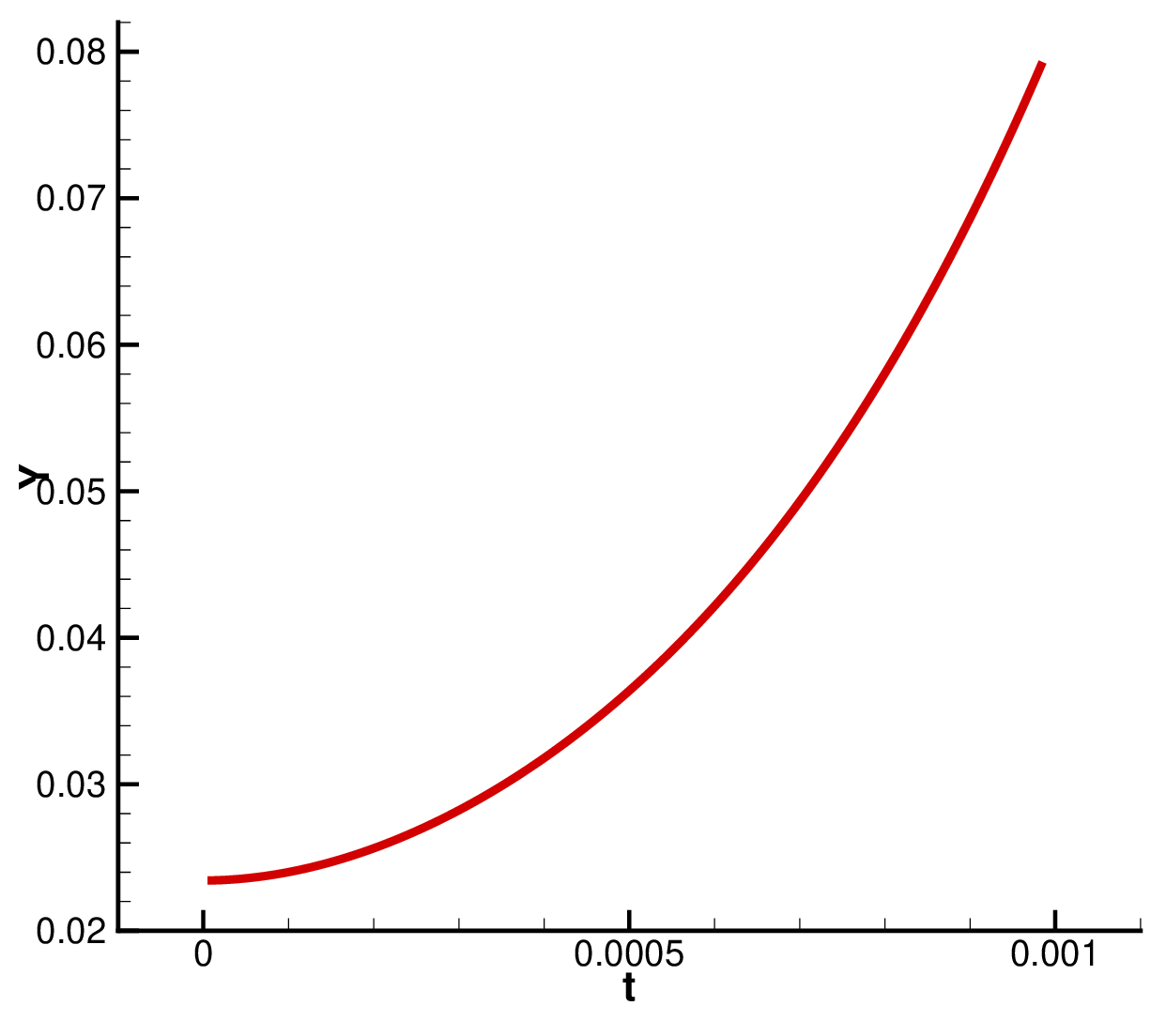}
    \caption{\label{fig:tstoY}}
  \end{subfigure}
  \begin{subfigure}{0.3\textwidth}
    \includegraphics[width=0.9\textwidth]{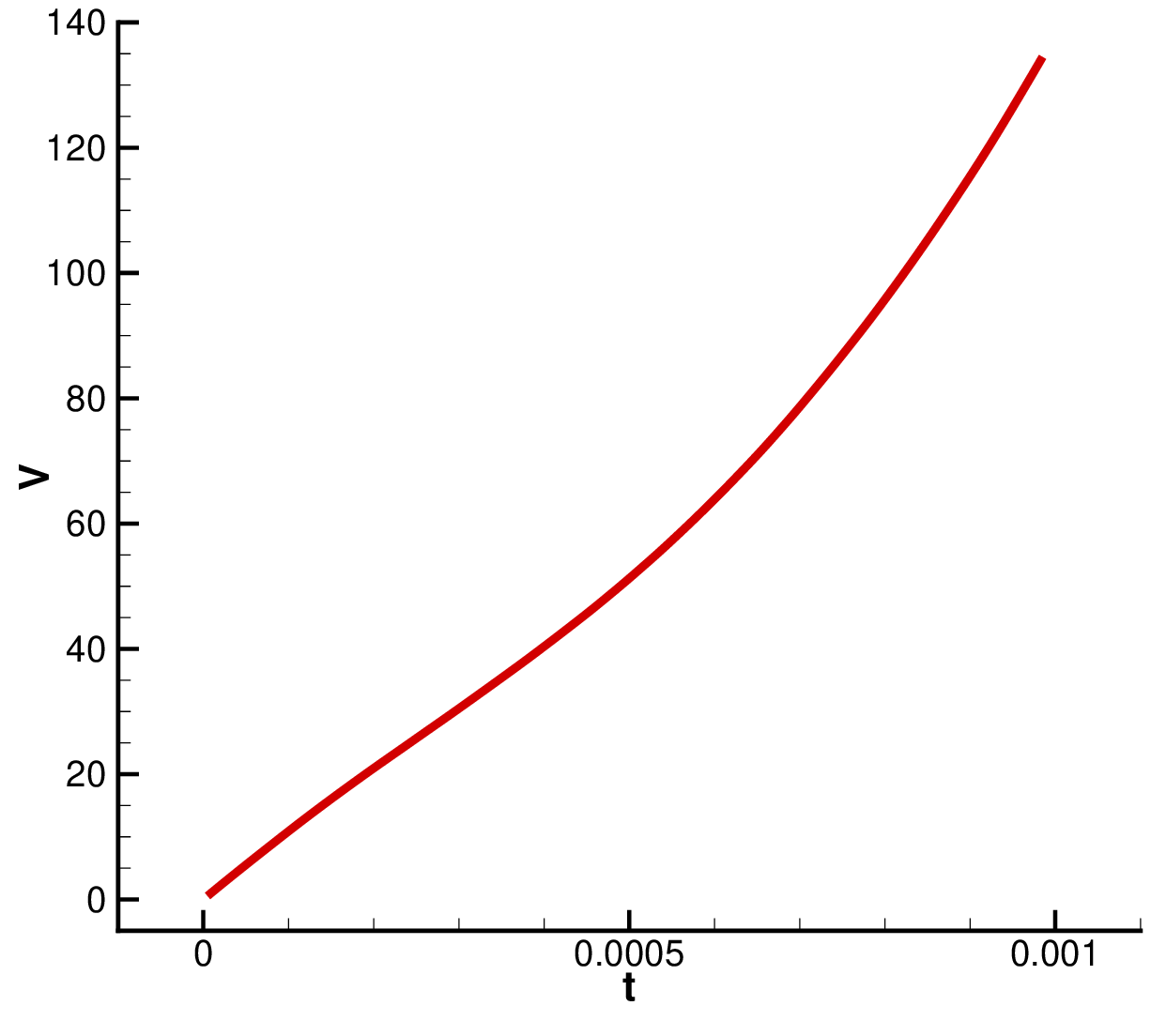}
    \caption{\label{fig:tstoV}}
  \end{subfigure}
  \begin{subfigure}{0.3\textwidth}
    \includegraphics[width=0.9\textwidth]{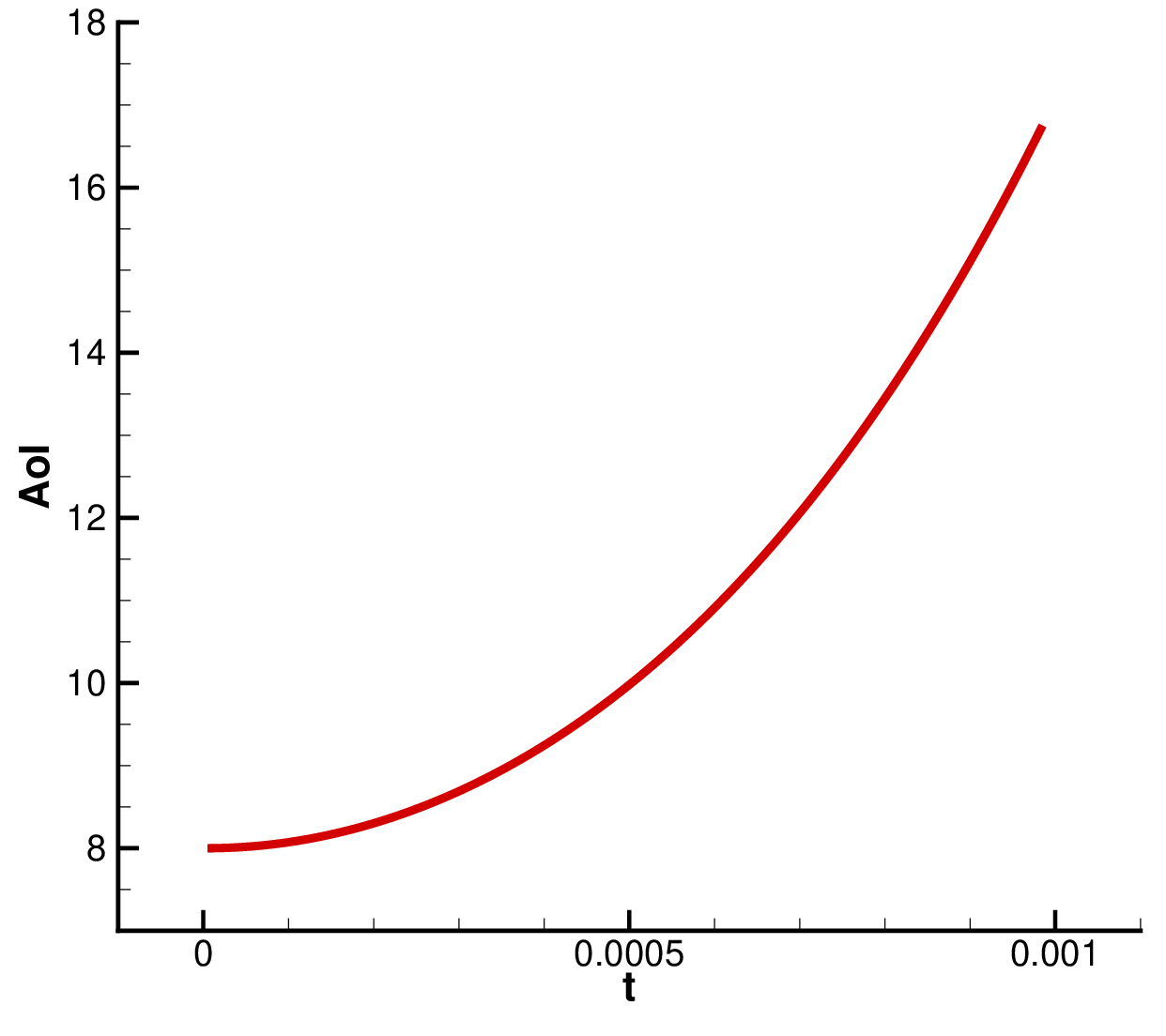}
    \caption{\label{fig:tstoAoI}}
  \end{subfigure}
  \begin{subfigure}{0.3\textwidth}
    \includegraphics[width=0.9\textwidth]{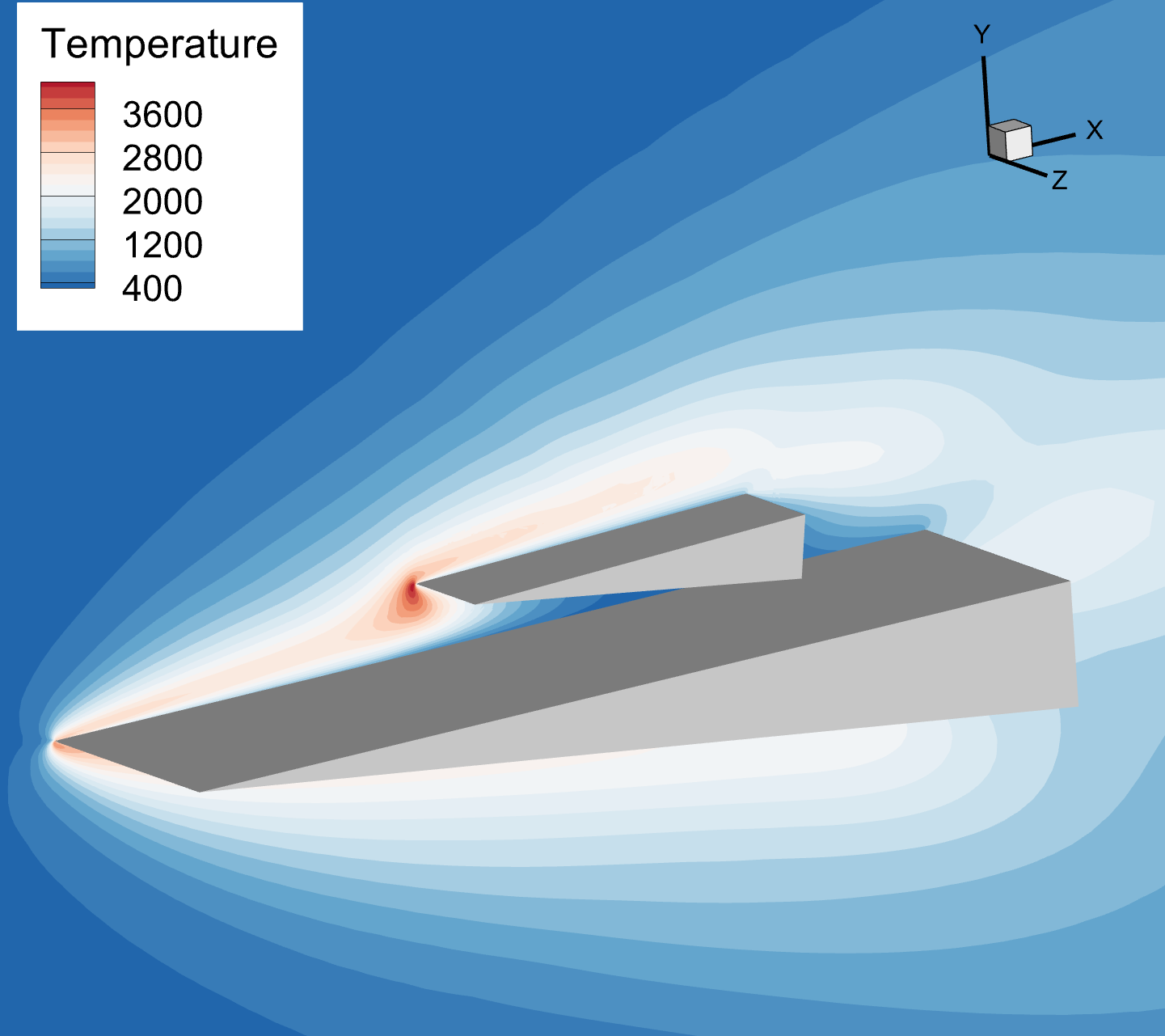}
    \caption{\label{fig:tsto0}}
  \end{subfigure}
  \begin{subfigure}{0.3\textwidth}
    \includegraphics[width=0.9\textwidth]{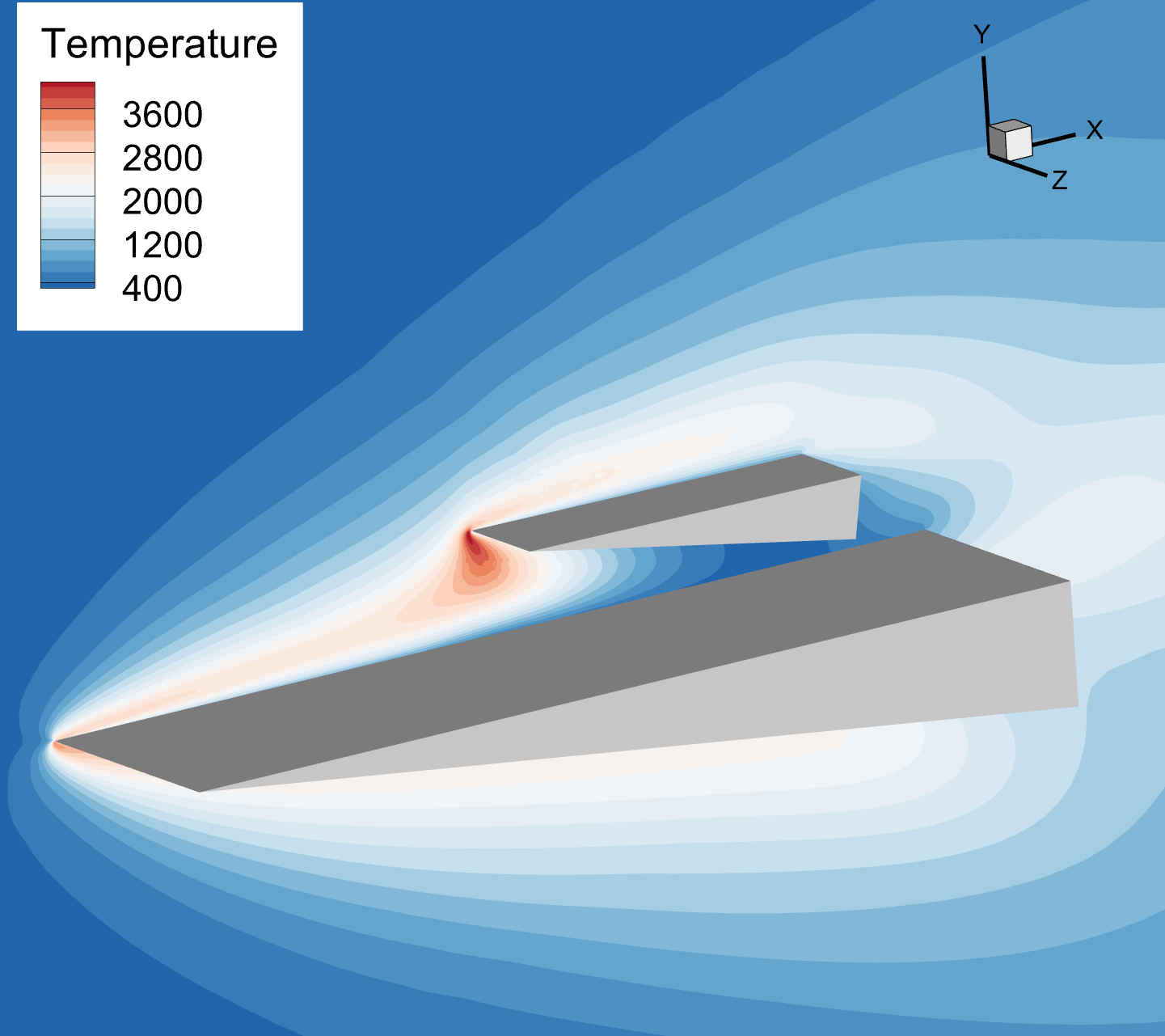}
    \caption{\label{fig:tsto5e-4}}
  \end{subfigure}
  \begin{subfigure}{0.3\textwidth}
    \includegraphics[width=0.9\textwidth]{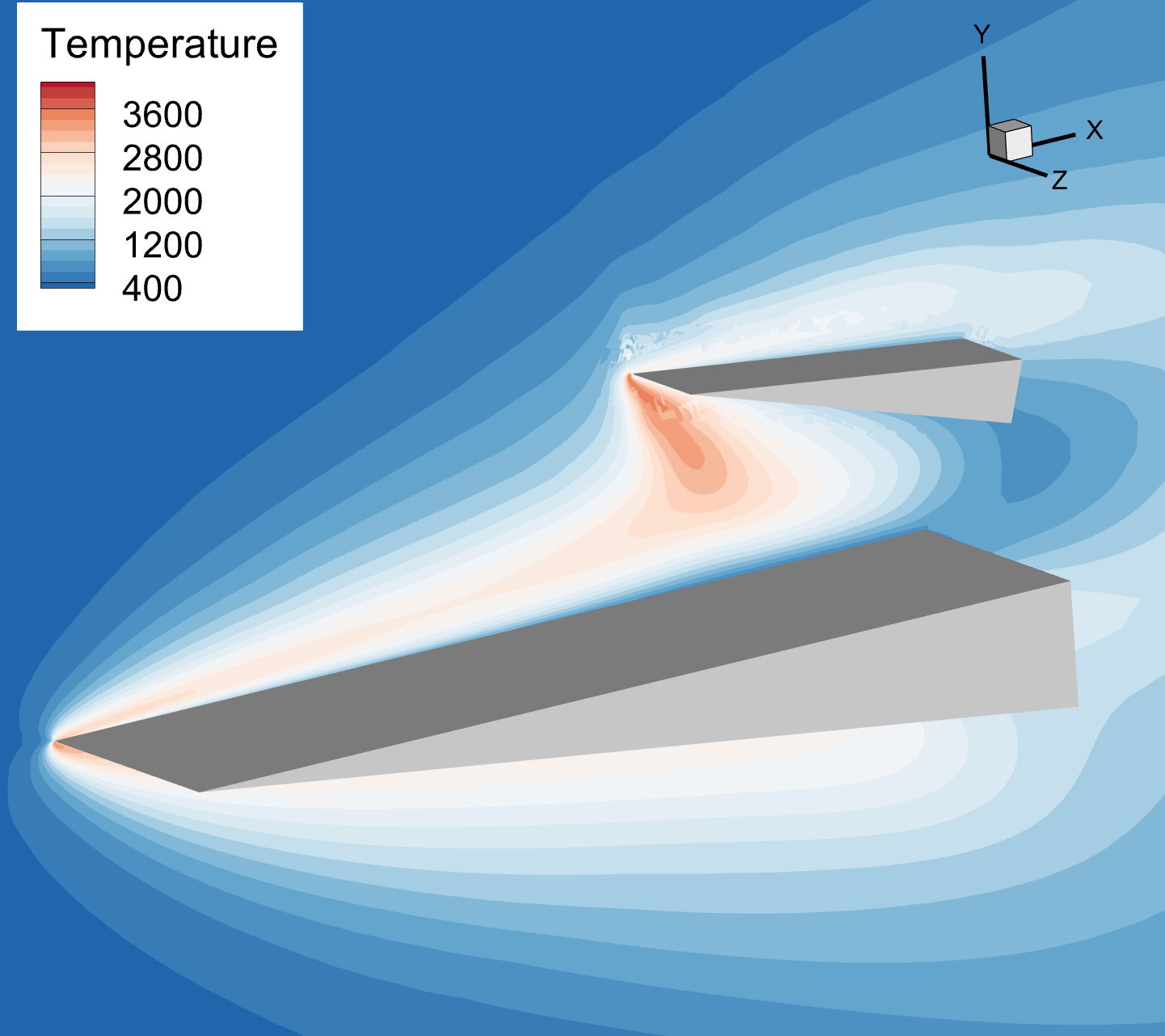}
    \caption{\label{fig:tsto9e-4}}
  \end{subfigure}
  \caption{The computational results of the two-stage-to-orbit (TSTO) hypersonic vehicles (a)The history position of orbiter's center of mass in the $y$-direction (b)The history velocity of orbiter's center of mass in the $y$-direction (c)The history angle of incidence. The pressure contours at different times (d) $t=0$ (e) $t=5\times 10^{-4} \text{s}$ (f) $t=9\times 10^{-4} \text{s}$}\label{fig:tstoResult}
\end{figure}

The computational results are shown in Fig. \ref{fig:tstoResult}.
The evolution of the orbiter's center of mass position in the $y$-direction, as shown in Fig. \ref{fig:tstoY}, exhibits a quadratic trend over time. As observed from the history of the $y$-direction velocity in Fig. \ref{fig:tstoV}, the acceleration first decreases and then increases. This is attributed to the initial reduction caused by the orbiter's rearward motion, followed by an increase in acceleration in the $y$-direction due to the subsequent increase in the angle of incidence.
The contour plots in Fig. \ref{fig:tsto0}, \ref{fig:tsto5e-4}, and \ref{fig:tsto9e-4} illustrate how changes in the relative positions of the two bodies lead to corresponding changes in the pressure distributions. The motion states show that the orbiter spits out of the booster in $y$-direction.

\section{Conclusion}

In this work, moving-mesh techniques and an overlapping-mesh assembly method are integrated into the Unified Gas-Kinetic Scheme (UGKS) to enable the efficient simulation of complex unsteady flows with moving boundaries. To maximize computational performance, the solver employs a high-performance programming paradigm coupled with a dual time-stepping algorithm, which significantly accelerates the unsteady simulations. The validity and robustness of the proposed method are demonstrated through a comprehensive series of test cases, ranging from two-dimensional micro-scale flows to complex three-dimensional vehicle separation problems. Ultimately, the numerical results confirm that the present approach achieves both high accuracy and exceptional computational efficiency when simulating challenging multiscale gas dynamics involving moving geometries.

\section*{Acknowledgements}
This work was supported by the National Natural Science Foundation of China (Grant No. 92371107), the National Key R$\&$D Program of China (Grant No. 2022YFA1004500), and the Hong Kong Research Grants Council (Grant No. 16208324).



\bibliographystyle{elsarticle-num}
\bibliography{sample}
\end{document}